\documentclass[twocolumn]{autart}
\usepackage{amsmath,amssymb}
\usepackage{mathrsfs}
\usepackage[dvips]{graphicx}
\usepackage{epsfig,psfrag} 
\usepackage{subfigure}
\usepackage{fancyhdr}
\usepackage{color}
\usepackage{verbatim}
\usepackage{natbib} % required for bibliography
\usepackage{algorithm}        
\usepackage{hyperref}
\usepackage{enumerate}

\newtheorem{theorem}{Theorem}[section]

\newtheorem{lemma}[theorem]{Lemma}

\newtheorem{definition}{Definition}[section]

\newtheorem{remark}{Remark}[section]
\newtheorem{problem}{Problem}[section]
\theoremstyle{definition}
\newtheorem{example}{Example}[section]

\newcommand{\RR}{{\mathbb R}}
\newcommand{\TT}{{\mathbb T}}

\newcommand{\cA}{{\mathcal A}}
\newcommand{\cB}{{\mathcal B}}

\newcommand{\cO}{{\mathcal O}}

\newcommand{\cS}{{\mathcal S}}

\newcommand{\cV}{{\mathcal V}}

\newcommand{\cY}{{\mathcal Y}}

\newcommand{\ol}{\overline}

\renewcommand{\Im}{\textup{Im }}

\newcommand{\aff}{\textup{aff }}

\newcommand{\conv}{\textup{co }}

\newcommand{\spn}{\textup{sp }}

\newcommand{\rank}{\textup{rank}}

\newcommand\xqed[1]{%
  \leavevmode\unskip\penalty9999 \hbox{}\nobreak\hfill
  \quad\hbox{#1}}
\newcommand\demo{\xqed{$\triangleleft$}}

\begin{document}

\begin{frontmatter}
%\title{Relaxed In-Block Controllability of Affine Systems on Polytopes }
\title{\bf On the Construction of Safe Controllable Regions for Affine Systems with Applications to Robotics } 
\thanks[footnoteinfo]{The authors are with the Dynamic Systems Lab (www.dynsyslab.org), Institute for Aerospace Studies, University of Toronto, Canada (e-mails: mohamed.helwa@robotics.utias.utoronto.ca, schoellig@utias.utoronto.ca). This research was supported by NSERC grant RGPIN-2014-04634 and the Connaught New Researcher Award.}%
\author{Mohamed K. Helwa},
\author{Angela P. Schoellig}%\ead{broucke@control.utoronto.ca}
%\address[Toronto]{Dept. of Electrical and Computer Engineering, University of Toronto, Toronto, ON M5S 3G4, Canada}

%\title{\bf On the Construction of Safe Controllable Regions for Affine Systems with Applications to Robotics 
%\thanks{The authors are with the Dynamic Systems Lab (www.dynsyslab.org), Institute
%for Aerospace Studies, University of Toronto, Canada (e-mails:
%mohamed.helwa@robotics.utias.utoronto.ca, schoellig@utias.utoronto.ca).
%This research was supported by NSERC grant RGPIN-2014-04634 and the Connaught New Researcher Award.}} 
%%Supported by theNatural Sciences and Engineering Research Council of Canada (NSERC).
%\author{Mohamed~K.~Helwa, and Angela P. Schoellig}% <-this % stops a space

\begin{abstract}
This paper studies the problem of constructing in-block controllable (IBC) regions for affine systems. That is, we are concerned with constructing regions in the state space of affine systems such that all the states in the interior of the region are mutually accessible through the region's interior by applying uniformly bounded inputs. We first show that existing results for checking in-block controllability on given polytopic regions cannot be easily extended to address the question of constructing IBC regions. We then explore the geometry of the problem to provide a computationally efficient algorithm for constructing IBC regions. We also prove the soundness of the algorithm. We then use the proposed algorithm to construct safe speed profiles for different robotic systems, including fully-actuated robots, ground robots modeled as unicycles with acceleration limits, and unmanned aerial vehicles (UAVs). Finally, we present several experimental results on UAVs to verify the effectiveness of the proposed algorithm. For instance, we use the proposed algorithm for real-time collision avoidance for UAVs.  

%This paper initiates the study of the problem of constructing in-block controllable covers of nonlinear systems on polytopes, which is useful in studying approximate mutual accessibility problems of nonlinear systems on polytopes. In particular, for a given nonlinear system and a given polytope, we study how to systematically construct polytopic covers of the given polytope such that the affine systems, resulting from the linearization of the nonlinear system inside the polytopic regions of the cover, are in-block controllable. By exploring the geometry of the problem, we first provide some constructive guidelines for building such covers. Then, for a special class of nonlinear systems, we provide a constructive algorithm. Illustrative examples of the main results are also provided.   
\end{abstract}

\end{frontmatter}

\section{Introduction}
In this paper, we introduce the problem of constructing in-block controllable (IBC) regions for affine systems. In particular, we study the construction of regions in the state space of affine systems such that all the states in the interior of the region are mutually accessible through the region's interior by applying uniformly bounded control inputs. We then use the proposed theoretical results to build safe speed profiles for several classes of robotic systems, including fully-actuated robot arms, ground robots with acceleration limits, and unmanned aerial vehicles (UAVs).

With the rapidly increasing desire for building the next generation of engineering systems that can safely interact with each other, their environment and possibly non-professional humans (e.g., self-driving cars or assistive robots), there is an urgent need for developing controller design methods that consider and obey to all given safety constraints of the systems even in the transient period. Hence, we set our goal to provide the mathematical foundations for controller design under safety constraints. 

There are two common ways for dealing with safety constraints in industrial control systems. First, typically, given safety constraints are not explicitly considered in the controller design phase but indirectly accounted for when manually tuning the controller parameters on the real system (e.g., PID controller design). Usually, an emergency system is added to the controller to shut down the industrial system in case of constraint violation (e.g., a robotic arm moving into an obstacle). However, shutting down the system is not possible in all applications, and considering the safety constraints in the design phase will prevent system damage, unnecessary system downtime, and therefore save money and time. Second, predictive/optimal controllers have received special interest for decades since they optimize the system's behavior, while respecting given, hard safety constraints \citep{predictive1,predictive2}. This is typically carried out by solving an online optimization problem within each sampling interval. Such predictive controllers have been successful in practice because of their ability to explicitly consider system constraints \citep{Qin2003}. Nevertheless, there are many fundamental questions in the area of controller design under safety constraints that still require further studies. For instance, suppose that we have a wheeled robot moving on a bounded table, with additional limits on the robot's speed. Using Kalman's controllability notion \citep{Kalman}, we cannot even answer the simple question whether the robot can reach, starting from any initial position and speed,  any final position and speed while respecting the safety constraint of staying on the table and using uniformly bounded input force?
%: starting from any initial position and speed of the robot, can we reach any final position and speed of the robot while respecting the safety constraints of staying on the table and using uniformly bounded input force? 
This illustrates the urgent need for finding checkable conditions that define when we can fully control our system within given safety constraints.

Hence, we recently introduced the study of in-block controllability (IBC), which formalizes Kalman's controllability under given safety state constraints \citep{HC14,HC16}. 
%There are other motivations behind the IBC notion that can be summarized as follows. 
The notion of IBC can, however, be motivated from several different perspectives.
In \citep{HC14_2}, we showed that if one constructs a special partition of the state space of piecewise affine (PWA) hybrid systems such that each region of the partition satisfies the IBC property, then one can systematically study controllability and build hierarchical structures for the PWA hybrid systems. We note that controllability of PWA hybrid systems is a challenging open problem to date \citep{CAMLIBEL1,CAMLIBEL2}. Also, building hierarchical structures of PWA hybrid systems allows us to design controllers that achieve temporal logic statements at the higher-levels of the hierarchy, and then to systematically realize these high-level control decisions at the lower levels. In \citep{HC15}, the IBC notion was also used to build special covers of the state space of nonlinear systems, in which each region satisfies the IBC property, and then these IBC covers were used to build hierarchical structures and to systematically study approximate mutual accessibility problems of nonlinear systems under safety constraints. Moreover, the IBC notion is useful in the context of optimal control problems. In particular,  the IBC conditions ensure that all the optimal accessibility problems within given safety constraints are feasible. Furthermore, in this paper we use the IBC results to build safe speed profiles for different classes of robotic systems. We then, for example, utilize these safe speed profiles to achieve static/dynamic obstacle avoidance. We also use the speed profiles to determine the feasibility of given reference trajectories, where we determine whether these reference trajectories are reachable from all other states in the safe position-speed region via trajectories that completely lie in the safe region and with inputs within the actuation limits.          

The notion of IBC was first introduced for finite state machines in \citep{Caines95}. The notion was then extended to nonlinear systems on closed sets in \citep{Caines98}, and to automata in \citep{Caines2002_2}. In these papers, the notion was used to build hierarchical control structures of the systems. However, these papers do not study conditions for when the IBC property holds. In \citep{HC14,HC16}, three necessary and sufficient conditions were provided for IBC of affine systems on given polytopes. The conditions require solving linear programming (LP) problems at the vertices of the given polytope. In \citep{H15}, the IBC conditions were extended to controlled switched linear systems having both continuous inputs and on/off control switches. In \citep{HC14_3,HC16_2}, the notion of IBC was relaxed to the case where one can distinguish between soft and hard safety constraints. Similar controllability studies to IBC can be found in \citep{Brammer}, \citep{Sontag}, \citep{Heemels2}, \citep{Heemels3}. In \citep{Brammer}, \citep{Sontag}, controllability of linear systems under input constraints was studied, while in \citep{Heemels2} controllability of continuous-time linear systems under state and/or input constraints was studied under the assumption that the system transfer matrix is right invertible. Under the same assumption, the study of \citep{Heemels2} was extended in \citep{Heemels3} to null controllability of discrete-time linear systems under constraints. Compared to the well-known controlled invariance problem \citep{blanchini,Hennet}, which requires that all the state trajectories initiated in a set to remain in the set for all future time, IBC has the additional requirement of achieving mutual accessibility. This is a basic, additional property that enables us to use IBC as a basis for building hierarchical control structures and for studying constrained mutual accessibility problems for PWA hybrid systems and nonlinear systems  \citep{HC14_2,HC15}. 

In many practical scenarios, however, it may happen that the given affine system is not IBC with respect to (w.r.t.) the given polytope, representing the intersection of the given safety constraints. For this case, it would be important from a practical perspective to find the largest IBC region inside the given region, formed by the intersection of the safety constraints. The IBC region then represents a large, safe region within which we can fully control our system. Also, constructing IBC regions is an essential problem for building the partitions/covers in \citep{HC14_2}, \citep{HC15}, respectively, so that one can make use of the hierarchical control results of these papers. This motivates us to study the problem of constructing IBC regions in this paper. 

In this paper, we first show the difficulties that are faced when trying to directly use the available results for checking IBC of affine systems on given polytopes to construct IBC regions. In particular, while checking the IBC property requires solving LP problems, building polytopes for which the IBC property holds generally requires solving bilinear matrix inequalities (BMIs), which is NP hard (see \citep{NP}). Second, we explore the geometry of the problem, and try to provide a computationally efficient method for constructing IBC polytopic regions, which avoids solving BMIs. Our geometric approach was first introduced in \citep{HC15_2} for a special class of affine systems, namely hypersurface systems for which $m=n-1$, where $m$ is the number of inputs and $n$ is the system dimension. In this paper, we extend the geometric study of \citep{HC15_2} to a more general geometric case that can be achieved for systems with $m\geq \frac{n}{2}$. We also provide a computationally efficient algorithm for constructing IBC polytopic regions, and prove its soundness. For our geometric study of IBC, we utilize some geometric tools that are used for the study of the control-to-facet problem, also called the reach control problems (RCP) on polytopes, \citep{HVS04, MEB10, HB13, HB15}. Third, we show how our proposed algorithm for constructing IBC regions can be useful for constructing safe speed profiles for different classes of robotic systems that include fully-actuated robots, ground robots modeled as unicycles with acceleration limits and UAVs. That is, we construct for each position of the robot a corresponding safe speed range. The proposed safe speed profiles are useful for robot speed scheduling algorithms \citep{Angela2014,Andrea,Ezquerro,Zucker}. In particular, if the speed scheduling algorithms limit the selected speeds to the safe speed profiles provided by our algorithm, then safety of the robot can be always achieved on the given constrained position space by applying a feasible input within the robot's actuation limits. We also show in this paper how the proposed safe speed profiles can be used to achieve static/dynamic obstacle avoidance. Compared to the safe speed profiles built by intuition or by the controlled invariance property, one advantage of the proposed safe speed profiles is that they guarantee full controllability of the robots on the position-speed regions constructed using our proposed algorithm. Therefore, there is no loss of generality in restricting the robots to operate in these constructed safe regions. As another advantage, the proposed algorithm ensures that any state in the constructed safe position-speed region is reachable from all other states in the safe region with trajectories that completely lie in the safe region and using a feasible input within the robot's actuation limits. Thus, in planning reference trajectories for robots, it would be important to select reference points inside the proposed safe position-speed regions to ensure that they are reachable within the given safety state constraints and under the robot's actuation limits. Compared to the feasibility study of \citep{Angela2}, we hereby take the safety position/speed constraints into consideration in determining the feasibility of given references, and not only the robot actuation limits.. Finally, in this paper we provide several experimental results on UAVs to verify the effectiveness of our proposed safe speed profiles.

The paper is organized as follows. Section \ref{sec:back} provides some geometric background. Section \ref{sec:IBC} reviews IBC. In Section \ref{sec:construct}, we introduce the problem of constructing IBC regions, provide a computationally efficient algorithm for solving the problem, and prove its soundness. In Section \ref{sec:rob}, we provide applications of the proposed algorithm to several classes of robotic systems. We summarize our results in Section \ref{sec:con}. A brief preliminary version of the paper appeared in \citep{HC16_3}. Here we include complete proofs, additional discussions and remarks, simulation results for fully-actuated and ground robots, and a novel, detailed subsection on UAVs, which includes experimental results.             

Notation:  
Let $K \subset \RR^n$ be a set. 
The closure of $K$ is denoted by $\ol{K}$, the interior by $K^{\circ}$, and the boundary by $\partial K$.  
%The notation $K_1\setminus K_2$ denotes the elements of the set $K_1$ not in $K_2$.
For vectors $x,~y\in \RR^n$, $x\cdot y$ denotes the inner product of the two vectors.
The notation $\left\|x\right\|$ denotes the Euclidean norm of $x$.
%The notation $\Zero$ denotes the subset of $\RR^n$ containing only the zero vector.
The notation $\conv\{ v_1,v_2,\ldots \}$ denotes the convex hull of a set of points $v_i \in \RR^n$.%, while $\aff\{ v_1,v_2,\ldots \}$ denotes the affine hull of a set of points $v_i \in \RR^n$.
%Finally, the $C^1$ stands for continuously differentiable. %, and $L_f V(x)$ is the Lie derivative of function $V :\RR^n \rightarrow \RR$ with respect to 
%function $f : \RR^n \rightarrow \RR^n$.
%The notation $\aff\{ v_1,v_2,\ldots \}$ denotes the affine hull of a set of points $v_i \in \RR^n$.
%Finally, $B_{\delta}(x)$ denotes the open ball of radius $\delta$ centered at $x$. 
%$T_{\cS}(x)$ denotes the Bouligand tangent cone to set $\cS$ at a point $x$. 
\section{Background}
\label{sec:back}
\begin{figure}[t]
\begin{center}
\includegraphics[scale=.27, trim = 0mm 95mm 10mm 45mm]{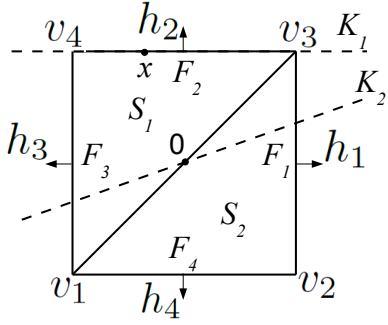} 
\end{center}
\caption{\scriptsize{Illustrative figure explaining several geometric concepts.}}
\label{fig_notation}
\end{figure}

We present some geometric background relevant for the remainder of the paper, see \citep{Brondsted2,ROCK}. 
A set $K\subset \RR^n$ is \emph{affine} if $\lambda x+(1-\lambda)y\in K$ for all $x,~y\in K$ and all $\lambda\in \RR$. An example are dashed, infinite lines $K_1$, $K_2$ in Figure \ref{fig_notation}. If the affine set passes through the origin, then it forms a subspace of $\RR^n$. For example, $K_2$ is a subspace of $\RR^2$. For subspaces $\cA,~\cB$, $\cA+\cB:=\{a+b~:~a\in \cA,~b\in \cB\}$. The set $\cA+\cB$ is also a subspace. The \emph{affine hull} of a set $K$, denoted by $\aff(K)$, is the smallest affine set containing $K$. We mean by a dimension of a set $K$ its affine dimension, which is the dimension of $\aff(K)$. For instance, in Figure \ref{fig_notation}, the dimension of $\conv\{v_3,v_4\}$ is the dimension of the affine set $K_1$, which is one. A \emph{hyperplane} is an $(n-1)$-dimensional affine set in $\RR^n$, dividing $\RR^n$ into two open half-spaces (e.g., $K_1$ in $\RR^2$). A finite set of vectors $\left\{x_1,\cdots,x_k\right\}$ is called \emph{affinely independent} if the unique solution to $\sum_{i=1}^{k}\alpha_i x_i=0$ and $\sum_{i=1}^{k}\alpha_i =0$ is $\alpha_i=0$ for all $i=1,\cdots,k$. Affinely independent vectors do not all lie in a common hyperplane. In Figure \ref{fig_notation}, the points $\{v_1,v_3,v_4\}$ are affinely independent, while the points $\{v_3,x,v_4\}$ are not. 
An \emph{$n$-dimensional simplex} is the convex hull of $(n+1)$ affinely independent points in $\RR^n$ (e.g., the triangles $S_1$ and $S_2$ in $\RR^2$). A simplex is a generalization of a triangle in 2D to arbitrary dimensions. An \emph{$n$-dimensional  polytope} is the convex hull of a finite set of points in $\RR^n$ whose affine hull has dimension $n$. Let $\left\{v_1,\cdots,v_p\right\}$ be a set of points in $\RR^n$, where $p>n$, and suppose that $\left\{v_1,\cdots,v_p\right\}$ contains (at least) $(n+1)$ affinely independent points. Then
$X:=\conv\left\{v_1,\cdots,v_p\right\}$ is an $n$-dimensional polytope. An example is the polytope $X=\conv\{v_1,\cdots,v_4\}$ in Figure \ref{fig_notation}.
A simplex is a special case of a polytope in which $p=n+1$. A \emph{face} of $X$ is any intersection of $X$ with a closed half-space such that none of the interior points of $X$ lie on the boundary of the half-space. The polytope $X$ and the empty set are considered trivial faces, and all other faces are called \emph{proper faces}. A \emph{facet} of $X$ is an $(n-1)$-dimensional face of $X$. A polytope is \emph{simplicial} if all its facets are simplices. We denote the facets of $X$ by $F_1,\cdots,F_r$, and we use $h_i$ to denote the unit normal vector to $F_i$ pointing outside of $X$. 
Figure \ref{fig_notation} illustrates this concept for $X = \conv\{v_1,\cdots,v_4\}$.
%See Figure \ref{fig_notation}.
%A family of $n$-dimensional sets $\left\{U_{i}~:~i\in Q\right\}$ is said to \emph{cover} $X$ if $X\subseteq \bigcup_{i\in Q}U_i$ \citep{Kelley}. By a polytopic cover, we mean a cover of $X$ consisting of sets, whose closures are polytopes.
In Section \ref{sec:construct}, we use triangulations of polytopes, and so we review its definition.
 \begin{definition} [\citep{LEE}]
A \emph{triangulation} $\TT$ of an $n$-dimensional polytope $X$ is a finite collection of $n$-dimensional simplices $S_1,\cdots,S_L$ such that:

(i) $X=\bigcup_{i=1}^{L}S_i$;

(ii) for all $i,~j\in\left\{1,\cdots,L\right\}$ with $i\neq j$, the intersection $S_i\cap S_j$ is either empty or a common face of $S_i$ and $S_j$.
\end{definition}
For example, $\TT=\{S_1,S_2\}$ is a triangulation of the polytope $X$ in Figure \ref{fig_notation}. 

\section{In-Block Controllability} 
\label{sec:IBC}
In this section, we review in-block controllability (IBC). Consider the affine control system:
\begin{equation}
\label{eq:thesystem}
\dot x(t) =A x(t) + Bu(t) + a \,, \qquad x(t) \in \RR^n \,,
\end{equation}
where $A \in \mathbb{R}^{n \times n}$, $a \in \mathbb{R}^n$, 
$B \in \mathbb{R}^{n \times m}$, and $\rank(B)=m$.
%Recall that $X^{\circ}$ denotes the interior of $X$.
Throughout the paper, we assume that the input $u:[0,\infty)\rightarrow \RR^m$ is measurable and bounded on any compact time interval to ensure the existence and uniqueness of the solutions of \eqref{eq:thesystem} \citep{Filippov}.
Let $\phi(x_0,t,u)$ be the trajectory of \eqref{eq:thesystem} under a control law $u$, with initial condition $x_0$ and evaluated at time $t$. We first review the IBC notion (after \citep{Caines98}).
\begin{definition}[In-Block Controllability (IBC)] 
\label{prob0}
Consider the affine control system \eqref{eq:thesystem} defined on an $n$-dimensional polytope $X$. We say that \eqref{eq:thesystem} is \emph{in-block controllable (IBC)} w.r.t. $X$ if there exists an $M>0$ such that for all $x,~y\in X^{\circ}$, there exist $T\geq 0$ and a control input $u$ defined on $[0,T]$ such that (i) $\left\|u(t)\right\|\leq M$ and $\phi(x,t,u)\in X^{\circ}$ for all $t\in [0,T]$, and (ii) $\phi(x,T,u)=y$.    
\end{definition}

That is, the system is IBC w.r.t. the polytope $X$ if all the states in the interior of $X$ are mutually accessible through its interior using uniformly bounded inputs. 

We review below the main result on IBC. 
In \citep{HC14}, it was shown that for studying IBC we can always apply a coordinate shift, and assume without loss of generality (w.l.o.g.) that we study a linear system
\begin{equation}
\label{eq:thesystem2}
\dot{\tilde{x}}(t) =A \tilde{x}(t) + B\tilde{u}(t) \,
\end{equation}
on a new polytope $\tilde{X}$ with $0\in \tilde{X}^{\circ}$. 
For notational convenience and w.l.o.g., we will call $\tilde{X}$, $\tilde{x}$, and $\tilde{u}$ just $X$, $x$, and $u$, respectively, in the remainder of the paper.
Let $J:=\left\{1,\cdots,r\right\}$ be the set of indices of the facets of $X$, and $J(x):=\left\{j\in J~:~x\in F_j\right\}$ be the set of indices of the facets of $X$ in which $x$ is a point. We define the closed, convex \emph{tangent cone} to $X$ at $x$ as
$C(x) := \{ y \in \RR^n : h_j \cdot y \le 0, ~j \in J(x) \},$
where $h_j$ is the unit normal vector to $F_j$ pointing outside $X$. 
%Then the main result of \citep{HC14} is reviewed below.
\begin{theorem} [\citep{HC14}]
\label{thm:main_IBC}
Consider the system \eqref{eq:thesystem2} defined on an $n$-dimensional simplicial polytope $X$ satisfying $0\in X^{\circ}$. The system \eqref{eq:thesystem2} is IBC w.r.t. $X$ if and only if
%\begin{itemize}
\begin{enumerate}[(i)]
\item $(A,B)$ is controllable;
\item the so-called invariance conditions of $X$ are solvable (that is, for each vertex $v\in X$, there exists $u\in \RR^m$ such that $Av+Bu\in C(v)$);
\item the so-called backward invariance conditions of $X$ are solvable (that is, for each vertex $v\in X$, there exists $u\in \RR^m$ such that $-Av-Bu\in C(v)$). %~~~~~~~~~~~~~~~~~~~~~~~~~~\QED
\end{enumerate}
%\end{itemize} 
\end{theorem}

In \citep{HC14}, it was shown that conditions (i)-(iii) of Theorem~\ref{thm:main_IBC} are also necessary for IBC on non-simplicial polytopes. For given polytopes, both the invariance conditions and the backward invariance conditions can be easily checked by solving a linear programming (LP) problem for each vertex of the polytope. The invariance conditions and the backward invariance conditions should only be checked at the vertices of $X$ since solvability of these conditions at the vertices implies by a simple convexity argument that they are solvable at all boundary points of $X$ \citep{HVS04}. 
\begin{remark}
\label{rem:input_con}
The definition of IBC can be easily tailored to the case when we have both state and input constraints. Suppose $u\in U\subset \RR^m$, where $U$ is a polytope having $0\in U^{\circ}$. For this case, the system is IBC if every $x,~y\in X^{\circ}$ are mutually accessible through $X^{\circ}$ using control inputs $u\in U$. Similarly, the definitions of invariance and backward invariance conditions are adapted to restrict $u$ to lie in $U$. It can be shown that for these tailored definitions, conditions (i)-(iii) of Theorem \ref{thm:main_IBC} remain necessary for IBC. 
%We will state in the next section a reasonable assumption on the set $U$ under which the proof of the sufficiency of conditions (i)-(iii) in this case is similar to the one in Section V of \citep{HC14}.  
Also, the proof of the sufficiency of conditions (i)-(iii) in this case is similar to the one in Section~V of \citep{HC14} under the mild assumption on $U$ that for any $\bar{x}\in X$ satisfying $A\bar{x}\in \Im(B)$, the image of $B$, there exists a $\bar{u}\in U^{\circ}$ such that $A\bar{x}+B\bar{u}=0$. %Details are omitted for brevity.  
\end{remark}

\section{Construction of IBC Regions}
\label{sec:construct}
In this section, we study the problem of constructing IBC regions for affine systems. The motivation behind the study is as follows. First, in many practical scenarios, it may turn out that the given dynamical system is not IBC w.r.t. the given region resulting from the intersection of the given safety constraints. Hence, it would be important to find a large IBC region within the given safety constraints, which represents a large safe region within which we can fully control our dynamical system. Second, the problem of constructing IBC regions is an important milestone towards building the special partitions/covers in \citep{HC14_2}, \citep{HC15}, respectively, so that one can make use of the hierarchical control results in these papers. Third, we show in Section~\ref{sec:rob} of the paper how the proposed results on constructing IBC regions can be useful for building safe speed profiles for different classes of robotic systems. These safe speed profiles are then utilized to achieve safe operation of robots, e.g. static/dynamic obstacle avoidance, and to determine the feasibility of given reference trajectories, in the sense that they can be reached from any safe initial condition within the given safety position/speed constraints and under the robot's actuation limits. 

Following \citep{HC14}, we know that w.l.o.g. the problem of studying IBC of an affine system can be transformed to studying a linear system on a new polytope $X$ having $0\in X^{\circ}$. Thus, we consider a linear system \eqref{eq:thesystem2}. Given the necessity of condition (i) of Theorem \ref{thm:main_IBC} for IBC, in our study of constructing IBC regions, we assume w.l.o.g. that \eqref{eq:thesystem2} is controllable. We then construct around the origin an IBC polytopic region for \eqref{eq:thesystem2}.
\begin{problem}[Construction of IBC Polytopes]
\label{prob1}
Given a controllable linear system \eqref{eq:thesystem2}, construct a polytope $X$ such that $0\in X^{\circ}$ and \eqref{eq:thesystem2} is IBC w.r.t. $X$.
\end{problem} 

It can be easily shown that if \eqref{eq:thesystem2} is IBC w.r.t. the polytope $X$, then it is also IBC w.r.t. $\lambda X:=\{x\in \RR^n~:~x=\lambda y,~y\in X\}$, a $\lambda$-scaled version of $X$, for every $\lambda>0$. Moreover, if all the mutual accessibility problems on $X^{\circ}$ are achieved using uniformly bounded inputs satisfying $\|u\|\leq M$, then all the mutual accessibility problems on $(\lambda X)^{\circ}$ can be achieved using uniformly bounded inputs satisfying $\|u\|\leq \lambda M$. 

While checking IBC on given polytopes is easy and incorporates solving LP problems as mentioned in the previous section, building IBC polytopic regions is considerably more difficult. Theorem \ref{thm:main_IBC} suggests that we build around the origin simplicial polytopes satisfying both the invariance conditions and the backward invariance conditions. Two difficulties are faced here. First, to build a polytope $X$ satisfying the invariance conditions (similar argument holds for the backward invariance conditions), we would need to select the vertices of $X$, $v_i$, the unit normal vectors to the facets of $X$, $h_j$, and the control inputs at the vertices, $u_i$, such that $h_j\cdot (Av_i+Bu_i)\leq 0$, for all $j\in J(v_i)$. Since $h_j$, $v_i$, and $u_i$ are all unknowns in this case, we have a set of bilinear matrix inequalities (BMIs), the solving of which is in general NP-hard \citep{NP}. Second, even if one constructs a polytope $X$ around the origin satisfying both the invariance conditions and the backward invariance conditions, one still needs to verify that $X$ is simplicial since the proof of the sufficiency of Theorem \ref{thm:main_IBC} only holds for simplicial polytopes. 

One possible approach to face these difficulties is as follows. Since the BMIs can be solved offline, one can exploit available software packages for solving BMIs such as PENBMI \citep{BMI}. Another possible approach is to use trial-and-error. In particular, one first constructs a candidate simplicial polytope $X$, and then uses Theorem \ref{thm:main_IBC} to check whether the given system is IBC w.r.t. $X$. If it is not the case, then one should try another candidate polytope, and so on. It is clear that these two approaches are computationally expensive, and for the second approach, there is no guarantee that one will eventually find the IBC polytope. Instead, in this paper, we explore the geometry of the problem, and try to provide a computationally efficient algorithm for building IBC polytopes that avoids solving BMIs. We initiated this geometric study in \citep{HC15_2} for hypersurface systems with $m=n-1$, and here we extend the study of \citep{HC15_2} to a more general geometric case.

To that end, let $\cB:=\Im(B)$ be the image of $B$, and define the set of possible equilibria of \eqref{eq:thesystem2}:
\begin{equation}
\label{BASICS:O}
\cO := \{ ~ x \in \RR^n ~:~ A x \in \cB ~ \} \,.
\end{equation}
At any point in $\cO$, the vector field of \eqref{eq:thesystem2} can vanish by proper selection of the input $u$. Also, if $x_0\in \RR^n$ is an equilibrium point of \eqref{eq:thesystem2} under some input, then $x_0\in \cO$ \citep{MEB10}. It can be verified that $\cO$ is closed, affine, and its dimension is $m\leq \kappa \leq n$ \citep{HB13}. Notice that both $\cB$ and $\cO$ are properties of the system \eqref{eq:thesystem2}, and, as such, they can be calculated before constructing the polytope $X$. 

For the geometric case $\cO + \cB = \RR^n$, we provide a computationally efficient algorithm for constructing IBC polytopes. We now show that this geometric condition is more general than the condition $m=n-1$ considered in \citep{HC15_2}. If $m=n-1$, then the dimension of $\cO$ is $n-1\leq \kappa \leq n$ \citep{HB13}. If $\kappa=n$, then $\cO + \cB = \RR^n$ clearly holds. We then show that $\cO + \cB = \RR^n$ holds for the case when $\kappa=n-1$. We claim that $\cB$ is not subset of $\cO$. Otherwise, we have $Ax+Bu\in \cB \subset \cO$ for all $x\in \cO$, and so $\cO$ is an invariant set under any selection of the control input $u$, which contradicts controllability of \eqref{eq:thesystem2}. If $\cB$ is not subset of $\cO$, then we can identify a non-zero vector $b\in \cB$ such that $b\notin \cO$. Since $\kappa=n-1$, then clearly $\cO + \cB = \RR^n$. On the other hand, for the following linear system, $\cO + \cB = \RR^n$ holds, while $m<n-1$:
%\small
\begin{equation}
\label{eq:rel}
\dot{x}(t) =
\left[ 
\begin{array}{cccc}
0 & 0 & 0 & 0 \\
0 & 0 & 0 & 0 \\
1 & 0 & 1 & 1 \\ 
0 & 1 & 0 & 1
\end{array} 
\right] x(t) +
\left[ 
\begin{array}{rr}
1 & 0 \\ 0 & 1 \\ 0 & 0 \\ 0 & 0 \end{array} 
\right] u(t).
\end{equation}
%\normalsize
This shows that the geometric case considered in this paper is more general than the one studied in \citep{HC15_2}. Indeed, since the dimension of $\cB$ is $m$ and the dimension of $\cO$ is $m\leq \kappa \leq n$ \citep{HB13}, the condition $\cO + \cB = \RR^n$ may be achieved for systems having $m\geq \frac{n}{2}$ as in \eqref{eq:rel}, which is a significant relaxation of the condition of \citep{HC15_2}\footnote{We present at this link: \url{https://drive.google.com/open?id=0BzU_Qe9rHozZTE82eUU2c2V0MWs} 
other examples of $(A,B)$ pairs for which $\cO + \cB = \RR^n$ is achieved (The examples are generated by MATLAB's rand command and are
saved as .mat files). This includes an example of a large-scale system with $n=1000$ and $m=500$.}. Also, we found that studying the geometric case $\cO + \cB = \RR^n$ is general enough to consider different classes of robotic systems in Section \ref{sec:rob}, including fully-actuated robot arms, ground robots, and unmanned aerial vehicles. Finally, we consider the study of the geometric case $\cO + \cB = \RR^n$ in this paper as a milestone in studying the general case in the future. 

We start by reviewing two geometric results of \citep{HC15_2}.
\begin{lemma}[\citep{HC15_2}]
\label{lem1}
Consider the linear system \eqref{eq:thesystem2}. For any polytope $X$, if $v\in \cO$ is a vertex of $X$, then the invariance conditions and the backward invariance conditions of $X$ are solvable at $v$. %~~~~~~~~~~~~~~~~~~~~~~~~~~~~~~~~~~~~~~~~~~~~~~~~~~~~~~~~\QED
\end{lemma}
\begin{lemma}[\citep{HC15_2}]
\label{lem2}
Consider the linear system \eqref{eq:thesystem2}. For any polytope $X$, if $\cB\cap C^{\circ}(v)\neq \emptyset$ at a vertex $v$ of $X$, where $C^{\circ}(v)$ denotes the interior of $C(v)$, then the invariance conditions and the backward invariance conditions of $X$ are solvable at $v$. %~~~~~~~~~~~~~~~~~~~~~~~~~~~~~~~~~~~~~~~~~~~~~~~~~~~~~~~~ \QED
\end{lemma}

Since $\cB$ and $\cO$ are properties of the linear system and can be calculated before constructing the polytope $X$, Lemmas \ref{lem1} and \ref{lem2} suggest that we can construct the polytope $X$ such that the vertices of $X$ lie on $\cO$, or the subspace $\cB$ dips into the interior of the tangent cones to the constructed polytope $X$ at the vertices. This ensures that both the invariance conditions and the backward invariance conditions are solvable at the vertices of the constructed polytope $X$. However, as mentioned before, there is still the difficulty that the proof of the sufficiency of Theorem \ref{thm:main_IBC} was carried out in \citep{HC14} only for simplicial polytopes, and, consequently, Theorem \ref{thm:main_IBC} may not apply. An extension of Theorem \ref{thm:main_IBC} is needed.

In this paper, we first show that for a given controllable linear system \eqref{eq:thesystem2}, if the vertices of the polytope $X$ are such that either $v\in \cO$ or $\cB\cap C^{\circ}(v)\neq \emptyset$, then the system \eqref{eq:thesystem2} is IBC w.r.t. $X$. We then provide, under the geometric condition $\cO+\cB=\RR^n$, a computationally efficient algorithm for constructing a polytope $X$ around the origin such that the vertices of $X$ satisfy $v\in \cO$ or $\cB\cap C^{\circ}(v)\neq \emptyset$. We also prove the soundness of the algorithm.

\begin{theorem} 
\label{thm:main_paper1}
Consider a controllable linear system \eqref{eq:thesystem2} defined on an $n$-dimensional polytope $X$ satisfying $0\in X^{\circ}$. If for each vertex $v$ of $X$, either $v\in \cO$ or $\cB\cap C^{\circ}(v)\neq \emptyset$, then the system \eqref{eq:thesystem2} is IBC w.r.t. $X$.   
\end{theorem}
\begin{pf}
%We provide a sketch of the proof in this conference version of the paper for brevity. In particular, 
By assumption and from Lemmas \ref{lem1}, \ref{lem2}, both the invariance conditions and the backward invariance conditions are solvable at the vertices of $X$. Although the three conditions of Theorem \ref{thm:main_IBC} hold, the polytope $X$ in our case is not necessarily simplicial, and consequently we cannot exactly follow the same sufficiency proof as in \citep{HC14} for Theorem \ref{thm:main_IBC}. Indeed, the proof of Theorem \ref{thm:main_IBC} is divided into three parts. In the first part, the invariance conditions are used to construct a continuous piecewise linear (PWL) feedback law, and under the assumption that the polytope $X$ is simplicial, it is proved that all the trajectories initiated in $X^{\circ}$ eventually tend to $\cO$ through $X^{\circ}$, and reach close to $\cO$ in finite time. Then, in the second part, controllability of $(A,B)$ is used to construct a piecewise continuous control input that makes the trajectories initiated nearby $\cO$ slide along $\cO$ inside $X^{\circ}$ towards $0\in X^{\circ}$ in finite time. Third, using the backward invariance conditions and a similar argument to the first two parts, it is shown that one can steer the backward dynamical system $\dot{x}=-Ax-Bu$ from any state in $X^{\circ}$ to the origin in finite time through $X^{\circ}$ using uniformly bounded inputs. Equivalently, one can steer the system \eqref{eq:thesystem2} from the origin to any final state in $X^{\circ}$ in finite time through $X^{\circ}$ using uniformly bounded inputs. One can see that the assumption that $X$ is simplicial is used in \citep{HC14} only in the first part of the proof to show that all trajectories initiated in $X^{\circ}$ tend to $\cO$. As a result, our task is reduced to prove this part in our case for any polytope, not necessarily simplicial. The rest of the proof is similar to \citep{HC14}. The details of the proof are in the appendix.
\end{pf}
We now provide under the geometric condition $\cO+\cB=\RR^n$ a computationally efficient algorithm for constructing a polytope $X$ such that $0\in X^{\circ}$ and the vertices of $X$ satisfy $v\in \cO$ or $\cB\cap C^{\circ}(v)\neq \emptyset$, which implies from Theorem \ref{thm:main_paper1} that the given system is IBC w.r.t. $X$. The algorithm is presented in Algorithm \ref{alg_paper1}. We then prove the soundness of the algorithm.
\begin{algorithm}
\caption{Construction of IBC polytopes}
\label{alg_paper1}
~~\\
\textbf{Given:} A controllable linear system \eqref{eq:thesystem2} satisfying $\cO+\cB=\RR^n$; Suppose $\cB=\spn \{b_1,\cdots,b_m\}$, and $\{o_{m+1},\cdots,o_n\}$ are such that $o_k\in \cO$ for all $k=m+1,\cdots,n$ and $\RR^n=\spn\{b_1,\cdots,b_m,o_{m+1},\cdots,o_n\}$.\\
\textbf{Objective:} Construct an $n$-dimensional polytope $X$ such that $0\in X^{\circ}$ and the system \eqref{eq:thesystem2} is IBC w.r.t. $X$.\\  
\textbf{Steps:}
\begin{enumerate}
\item Construct an initial $n$-dimensional polytope $P$ such that $0\in P^{\circ}$, and let $\{v_1,\cdots,v_p\}$ denote the vertices of $P$. 
\item Let $T=[b_1~\cdots~b_m~o_{m+1}~\cdots~o_n]$ and $T_{\cO}=[0~\cdots 0~o_{m+1}~\cdots~o_n]$. For $v_i$, $i=1,\cdots,p$, calculate $\bar{o}_i=T_{\cO}T^{-1}v_i$. %\footnote{We emphasize that $T^{-1}$ is calculated only once.}.
\item Select $\alpha>1$, and define $\tilde{o}_i:=\alpha\bar{o}_i$ for $i=1,\cdots,p$. 
\item Define $X:=\conv\{v_1,\cdots,v_p,\tilde{o}_1,\cdots,\tilde{o}_p\}$. 
\end{enumerate} 
\end{algorithm}
\begin{theorem} 
\label{thm:main_paper2}
Consider a controllable linear system \eqref{eq:thesystem2} satisfying $\cO+\cB=\RR^n$. Then, Algorithm \ref{alg_paper1} always terminates successfully, and the system \eqref{eq:thesystem2} is IBC w.r.t. the constructed polytope $X$.  
\end{theorem}
\begin{pf}
Since $\cO+\cB=\RR^n$, one can always identify $o_{m+1},\cdots,o_{n}$ such that $o_k\in \cO$ for all $k=m+1,\cdots,n$, and $\RR^n=\spn\{b_1,\cdots,b_m,o_{m+1},\cdots,o_n\}$. Since $T$ has linearly independent columns, it is invertible. Hence, one can always calculate $\bar{o}_i, \tilde{o}_i$, and then construct $X$. By construction, $0\in P^{\circ}\subset X^{\circ}$. 

We now show that \eqref{eq:thesystem2} is IBC w.r.t. $X$. To that end, we prove that the vertices of $X$ satisfy $v\in \cO$ or $\cB\cap C^{\circ}(v)\neq \emptyset$. Notice that the vertices of $X$ are subset of $\{v_1,\cdots,v_p,\tilde{o}_1,\cdots,\tilde{o}_p\}$. Let $c_i=(c_{i1},c_{i2},\cdots,c_{in}):=T^{-1}v_i$. It is straightforward to show $v_i=\sum_{j=1}^{m}c_{ij}b_j+\sum_{j=m+1}^{n}c_{ij}o_j$, $\sum_{j=1}^{m}c_{ij}b_j=:b_{v_i}\in \cB$, and $\sum_{j=m+1}^{n}c_{ij}o_j\in \cO$. From step 2, $\bar{o}_i=T_{\cO}c_i=\sum_{j=m+1}^{n}c_{ij}o_j\in \cO$. Thus, we have 
\begin{equation}
\label{eq:pf_alg}
v_i=b_{v_i}+\bar{o}_i. 
\end{equation}
Since $\cO$ is affine and $0\in \cO$, $\tilde{o}_i:=\alpha \bar{o}_i \in \cO$. We then study the vertices of $X$ in the set $\{v_1,\cdots,v_p\}$. Notice that $\bar{o}_i\in \conv\{\tilde{o}_i,0\}$, and if $\bar{o}_i\neq 0$, then $\tilde{o}_i\neq \bar{o}_i$. Since $\tilde{o}_i\in X$ by construction and $0\in X^{\circ}$, then $\bar{o}_i\in X^{\circ}$. Now if $v_i$, $i\in\{1,\cdots,p\}$, is a vertex of $X$, then from \eqref{eq:pf_alg}, $v_i-b_{v_i}=\bar{o}_i\in X^{\circ}$, which implies that $-b_{v_i}\in \cB$ dips into the interior of the tangent cone to $X$ at $v_i$, i.e. $-b_{v_i}\in \cB\cap C^{\circ}(v_i)\neq \emptyset$. From Theorem \ref{thm:main_paper1}, \eqref{eq:thesystem2} is IBC w.r.t. $X$.     
\end{pf}

\begin{remark}
\label{rem_comp_eff}
Notice that in Step 2 of Algorithm \ref{alg_paper1}, $T^{-1}$ should be calculated only once. Indeed, Algorithm \ref{alg_paper1} does not require solving any optimization problem, which represents a significant reduction of computational complexity compared to the original formulation of the problem that requires solving BMIs or using trial-and-error. Computational efficiency is quite important in fast applications. For instance, in Section \ref{sec:rob}, we compute the IBC regions (the safe speed profiles) for UAVs at each sampling instant to avoid dynamic obstacles that intersect with the vehicle's path.    
\end{remark}

\begin{remark}
\label{rem_scal}
As discussed before, for any $\lambda>0$, \eqref{eq:thesystem2} is also IBC w.r.t. $\lambda X$ using $\lambda$-scaled inputs of the ones used to solve mutual accessibility problems on $X^{\circ}$. This may be useful in two ways. First, if it is required to keep the system within given, hard safety constraints that form a region $X_c$ around the origin, then one can first use Algorithm \ref{alg_paper1} to construct an IBC polytopic region $X$ satisfying $0\in X^{\circ}$, and then one can simply scale $X$ such that $\lambda X\subset X_c$. Here, $\lambda X$ represents a safe region, within which we can fully control our system. Second, for the case of input constraints ($u\in U\subset \RR^m$, where $0\in U^{\circ}$), we can similarly scale $X$ such that on $\lambda X$, $\lambda<1$, the IBC property is achieved using $u\in U$.
\end{remark}

We present a simple illustrative example of Algorithm \ref{alg_paper1}.
\begin{example}
\label{ex_alg}
Consider the double integrator $\dot{x}_1=x_2$, $\dot{x}_2=u$. The system is evidently controllable. We have $\cO=\{x\in \RR^2~:~x_2=0\}$, the $x_1$-axis, and $\cB=\spn\{(0,1)\}$, the $x_2$-axis. Hence, $\cO+\cB=\RR^2$. We follow the steps of Algorithm \ref{alg_paper1}: (1) We construct $P=\conv\{v_1,\cdots,v_4\}$, where $v_1=(-0.8,-1)$, $v_2=(0.8,-1)$, $v_3=(0.8,1)$, and $v_4=(-0.8,1)$\footnote{One can easily verify using Theorem \ref{thm:main_IBC} that the system is not IBC w.r.t. $P$.}; (2) we have $b_1=(0,1)$, $o_2=(1,0)$, and we calculate $\bar{o}_1=\bar{o}_4=(-0.8,0)$ and $\bar{o}_2=\bar{o}_3=(0.8,0)$; (3) we select $\alpha=1.25$, and so $\tilde{o}_1=\tilde{o}_4=(-1,0)$ and $\tilde{o}_2=\tilde{o}_3=(1,0)$; (4) the system is IBC w.r.t. $X=\conv\{v_1,\cdots,v_4,\tilde{o}_1,\tilde{o}_2\}$ shown in Figure~\ref{fig:ex_DI}. \demo
\begin{figure}[t]
\begin{center}
\includegraphics[scale=.24, trim = 10mm 95mm 10mm 15mm]{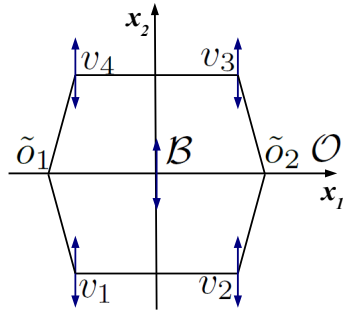} 
\end{center}
\caption{\scriptsize{The constructed IBC polytope $X$ in Example \ref{ex_alg}.}}
\label{fig:ex_DI}
\end{figure}
\end{example}
\section{Applications to Robotics}
\label{sec:rob}
In this section, we show how our proposed algorithm, Algorithm \ref{alg_paper1}, can be useful in constructing safe speed profiles for different robotic systems that include fully-actuated robot arms, ground robots, and unmanned aerial vehicles (UAVs). We also highlight the advantages of the proposed safe speed profiles compared to the ones obtained by intuition or by constructing controlled invariant sets in the position-speed state space. Moreover, in addition to the simulation results presented in this section, we present several experimental results on UAVs to verify the effectiveness of our proposed results. We start this section with fully-actuated robots. 
\subsection{Fully-Actuated Robots}
Consider a fully-actuated robot with $N$ links that is modeled by:
\begin{equation}
\label{eq:fully}
D(q)\ddot{q}+C(q,\dot{q})\dot{q}+g(q)=B(q) \tau,
\end{equation}
where $q=(q_1,\cdots,q_N)$ is the vector of generalized coordinates\footnote{The element $q_i$ represents the angle of link $i$ if joint $i$ is revolute (we assume $q_i\in (-\pi,\pi]$), or it is the displacement if joint $i$ is prismatic.}, $\dot{q}=(\dot{q}_1,\cdots,\dot{q}_N)$ is the vector of velocities, $\tau$ is the vector of generalized applied forces\footnote{That is forces and/or torques.}, and $D$ is a positive definite matrix. For fully-actuated robots, it is well-known that $B\in \RR^{N\times N}$ is full-rank, and so one can use the feedback law
\begin{equation}
\label{eq:fblin}
\tau=B^{-1}(q)(C(q,\dot{q})\dot{q}+g(q)+D(q)u)
\end{equation}
to convert \eqref{eq:fully} into the equivalent controllable linear system
\begin{equation}
\label{eq:lin_rob}
\ddot{q}=u,
\end{equation}
which is a set of decoupled double integrators, $\ddot{q}_i=u_i$, representing the dynamics in the different generalized coordinates. 

Suppose that we have the position constraints $q_i\in [q_{i,min},q_{i,max}]$, the velocity limits of the robot $\dot{q}_i\in [\dot{q}_{i,min},\dot{q}_{i,max}]$, where $0\in (\dot{q}_{i,min},\dot{q}_{i,max})$, and the actuator limits $\tau_i \in [\tau_{i,min},\tau_{i,max}]$, where $0\in (\tau_{i,min},\tau_{i,max})$. Assume that the position space is free of kinematic singularities, and that w.l.o.g. $0\in (q_{i,min},q_{i,max})$ for each $i$. Operating the robot within the maximum velocity limits does not ensure that the robot remains within the required position limits, and consequently does not ensure safety of operation such as collision avoidance. Instead, it is required to define a safe speed profile for the robot. That is, for each value of $q_i$ within the position limits, we define a corresponding range of safe velocities, resulting in an overall safe region in the position-velocity state space. 

We assume that for the given position-speed limits, the feedback linearization \eqref{eq:fblin} can be carried out within the actuator limits of the robot, provided that for each $i$, $u_i$ is selected within $[u_{i,min},u_{i,max}]$, where $0\in (u_{i,min},u_{i,max})$. Hence, our task is reduced to finding for the equivalent linear system \eqref{eq:lin_rob} a safe controllable region, within the given position-speed ranges, while taking into consideration the limits on the inputs $u_i$. It is straightforward to verify that for the controllable linear system \eqref{eq:lin_rob}, $\cO+\cB=\RR^{2N}$, and so Algorithm \ref{alg_paper1} can be used to find a controllable safe position-speed region. Indeed, since \eqref{eq:lin_rob} is a set of decoupled double integrators, one can apply Algorithm \ref{alg_paper1} for each subsystem $\ddot{q}_i=u_i$ to find a safe speed profile for each generalized coordinate $q_i$ (similar problem to Example \ref{ex_alg}). 

As discussed in Remark \ref{rem_scal}, although Algorithm \ref{alg_paper1} does not directly take the actuator limits into consideration in calculating the IBC polytope $X$, one can always scale the obtained polytope $X$ to find another IBC polytope $\lambda X$, in which all the mutual accessibility problems are achieved using control inputs within the actuator limits. For the double integrator example ($\dot{x}_1=x_2$, $\dot{x}_2=u$, $u\in [u_{min},u_{max}]$, where $0\in (u_{min},u_{max})$), this can be simply done as follows. One should first verify after constructing the IBC polytope $X$ using Algorithm \ref{alg_paper1} that at each vertex of $X$ not in $\cO=\{x\in \RR^2~:~x_2=0\}$, both the strict invariance conditions and the strict backward invariance conditions are achieved using inputs $u\in [u_{min},u_{max}]$. Since the polytope $X$ is known from Algorithm \ref{alg_paper1}, this verification can be carried out by solving LP problems. If the verification result is positive, then in spite of the actuator limits, we can still construct the special PWL feedback $u_p(x)$ in the proof of Theorem \ref{thm:main_paper1}, and it can be shown that the system is IBC w.r.t. $X$ using inputs $u$ satisfying $u\in [u_{min},u_{max}]$. Instead, if the verification result is negative, then with the aid of the fact that here $\cB=\spn\{(0,1)\}$, it can be shown that one can always scale the $x_2$-component of the vertices of $X$ (scale the velocity profile) to end up with a new IBC polytope $X'$ for which the mutual accessibility problems are achieved using inputs $u$ within $[u_{min},u_{max}]$. 

\begin{figure}[t]
\begin{center}
\includegraphics[scale=.23, trim = 10mm 105mm 10mm 15mm]{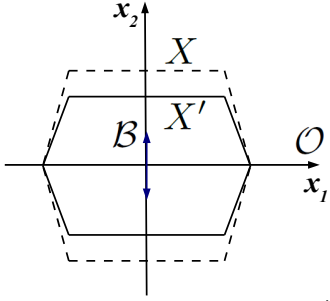} 
\end{center}
\caption{\scriptsize{The constructed IBC polytopes $X$, $X'$ for constraints $-10 \leq \tau \leq 10$ and $-8 \leq \tau \leq 8$, respectively.}}
\label{fig:ex_DI_lim}
\end{figure}

\begin{figure}[t]
\begin{center}
\includegraphics[scale=.32, trim = 10mm 10mm 10mm 15mm]{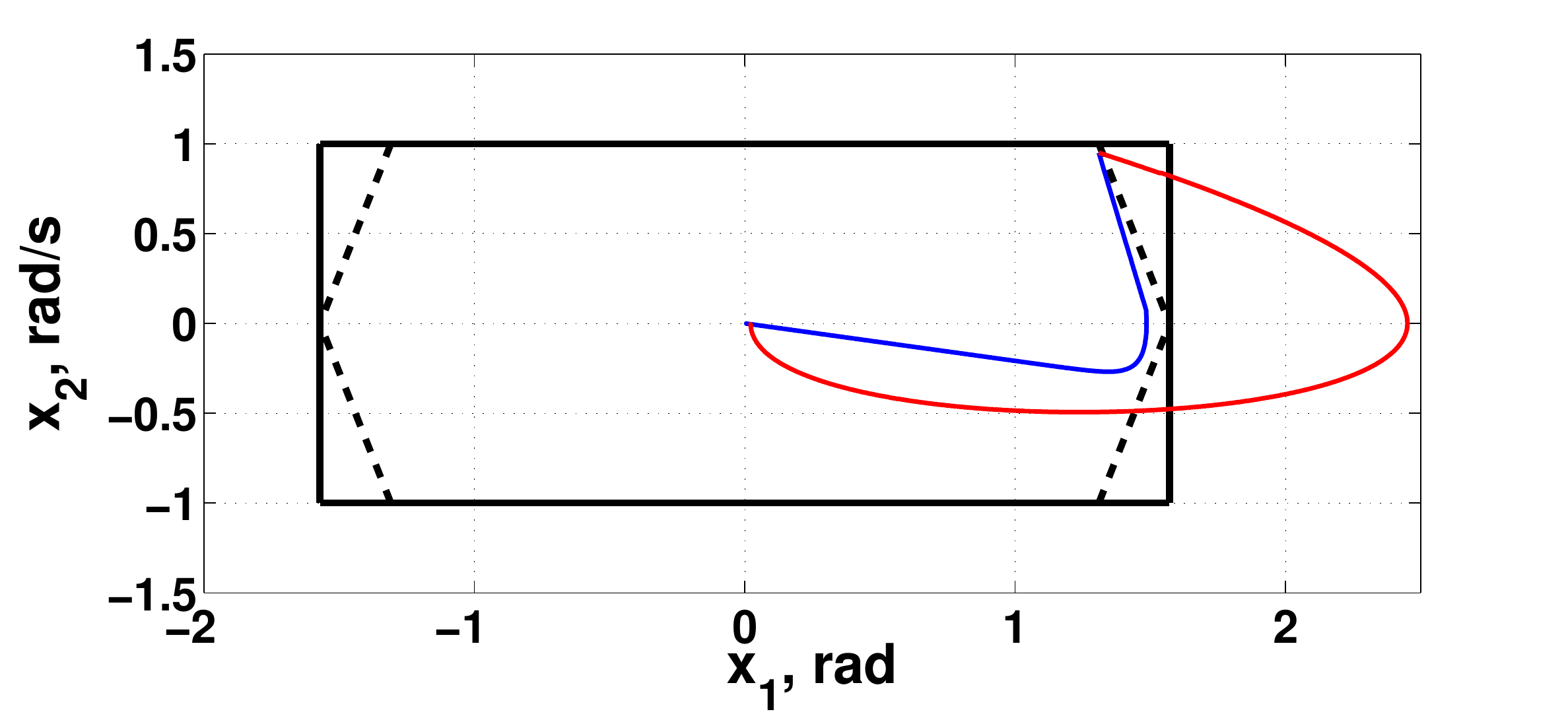} 
\end{center}
\caption{\scriptsize{An example of connecting two state points of the robot arm under the actuation limits $-10\leq \tau \leq 10$: the black dashed lines form the safe speed profile for the one DOF robot; the red trajectory is under the traditional control law in equation (15) of \citep{HC14}, with $t_f=10~s$; the blue trajectory is the proposed one.}}
\label{fig_oneDOF}
\end{figure}

To make our discussion more concrete, consider, for instance, a one degree-of-freedom (DOF) robot arm represented by

\begin{equation}
\label{eq:fully_ex}
I\ddot{\theta}=-mgl~sin(\theta)+\tau,
\end{equation}

where $\theta$ is the robot angle, $I$ is its inertia, $m$ is its mass, $l$ is the robot arm length, $g$ is the gravitational acceleration constant, and $\tau$ is the input torque. Suppose that $I=1~kg. m^2$, $m=1~kg$, $l=0.5~m$, and $g=10~ m/{s}^2$. Also, suppose that we have the state constraints $-\frac{\pi}{2}\leq \theta \leq \frac{\pi}{2}$, $-1\leq \dot{\theta}\leq 1$, and the input constraints $-10 \leq \tau \leq 10$. By using the feedback linearization law
\begin{equation}
\label{eq:fully_ex2}
\tau=mgl~sin(\theta) + Iu,
\end{equation} 
we get the linearized dynamics
\begin{equation}
\label{eq:fully_ex3}
\ddot{\theta}=u,
\end{equation}
which is a double integrator. It is straightforward to verify that if $-5\leq u \leq 5$, then with \eqref{eq:fully_ex2}, $-10\leq \tau \leq 10$, i.e. the actuator limits of the robot arm are satisfied. Hence, for \eqref{eq:fully_ex3}, it is required to find a safe controllable position-speed region under the limits $-5\leq u \leq 5$. Similar to Example \ref{ex_alg}, we use Algorithm \ref{alg_paper1} to construct the IBC polytope $X=\conv\{v_1,\cdots,v_4,\tilde{o}_1,\tilde{o}_2\}$ shown in Figure \ref{fig:ex_DI_lim}, where $v_1=(-\frac{5\pi}{12},-1)$, $v_2=(\frac{5\pi}{12},-1)$, $v_3=(\frac{5\pi}{12},1)$, $v_4=(-\frac{5\pi}{12},1)$, $\tilde{o}_1=(-\frac{\pi}{2},0)$ and $\tilde{o}_2=(\frac{\pi}{2},0)$. 
One can easily verify that under $-5\leq u \leq 5$, both the strict invariance conditions and the strict backward invariance conditions are solvable at the vertices outside $\cO$, and consequently the system is IBC w.r.t. $X$ under the given actuator limits. Now suppose that we have tighter actuator limits $-8 \leq \tau \leq 8$. For this case, $-3\leq u \leq 3$ ensures under \eqref{eq:fully_ex2} that the robot's actuator limits are satisfied. Then under $-3\leq u \leq 3$, it can be easily verified that the invariance conditions are not solvable at the vertex $v_3=(\frac{5\pi}{12},1)\notin \cO$. As a result, we should scale the set $X$, or as discussed above, scale the velocity-component ($x_2$-component) of the vertices not in $\cO$. For a scaling factor $\lambda=0.75$ of the velocity components, one can verify that for the new polytope $X'=\conv\{v_1',\cdots,v_4',\tilde{o}_1,\tilde{o}_2\}$ shown in Figure \ref{fig:ex_DI_lim}, where $v_1'=(-\frac{5\pi}{12},-0.75)$, $v_2'=(\frac{5\pi}{12},-0.75)$, $v_3'=(\frac{5\pi}{12},0.75)$, and $v_4'=(-\frac{5\pi}{12},0.75)$, both the strict invariance conditions and the strict backward invariance conditions are solvable at the vertices of $X'$ not in $\cO$ using control inputs that satisfy $-3\leq u\leq 3$. Hence, $X'$ satisfies the IBC property under $-8 \leq \tau \leq 8$. One can see that with $X$ or $X'$, we provide for each position within the given limits a corresponding safe speed range, staying within those guarantees us that the system is safe at all times. This safe profile can inform learning-based speed scheduling algorithms \citep{Angela2014}, which gradually increase a robot's speed based on information from previous runs. Using the same example, suppose that it is required under the actuator limits $-10\leq \tau \leq 10$ to connect the state point $x_0=(\theta_0,\dot{\theta}_0)=(\frac{5\pi}{12},0.95)$ to the origin $(0,0)$ in finite time within the given state constraints. Since both state points lie in the IBC region $X$, we know that we can find control inputs satisfying the constrained mutual accessibility in finite time under the given actuator limits. Figure \ref{fig_oneDOF} shows two trajectories connecting $x_0$ to the origin: the red trajectory is obtained by applying the traditional control law of connecting two states based on the control Gramian, equation (15) of \citep{HC14}, with $t_f=10~s$, while the blue trajectory is obtained by first using the PWL feedback discussed in the proof of Theorem \ref{thm:main_paper1} to decelerate the robot arm and avoid violating the safety state constraints, and then using a traditional control law. One can see that using the traditional control law (equation (15) of \citep{HC14}), there is no guarantee that the state constraints are satisfied in the transient phase.   

\begin{figure}[t]
\begin{center}
\includegraphics[scale=.2, trim = 10mm 120mm 10mm 15mm]{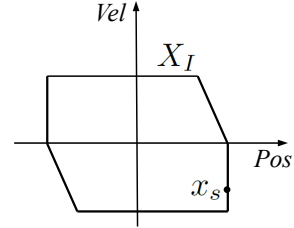} 
\end{center}
\caption{\scriptsize{A safe speed profile obtained by intuition.}}
\label{fig:ex_DI_Int}
\end{figure}

We now show the advantages of the proposed safe speed profiles compared to the ones obtained by intuition. One can simply argue that to prevent the violation of the position constraints near the edge $x_1=\frac{\pi}{2}$, only a reduced
forward velocity is allowed. Similarly, to prevent the violation of the position constraints near the edge $x_1=-\frac{\pi}{2}$, only a reduced backward velocity is allowed. This results in a polytope $X_I$ shown in Figure \ref{fig:ex_DI_Int}, which represents a safe speed profile obtained by intuition.  
Our proposed speed profiles in Figure \ref{fig:ex_DI_lim} have the following advantages compared to the intuitive one in Figure \ref{fig:ex_DI_Int}. First, our proposed method provides a systematic approach for obtaining the vertices of the polytopic safe region, which is an advantage compared to the intuitive method, especially for more complicated systems. Second, our constructed polytopes satisfy the IBC property, and consequently all the states in the interior of the polytope are mutually accessible through its interior by applying control inputs within the actuator limits. Thus, there is no loss of generality (in terms of controllability) in restricting the robot to operate in the proposed safe position-speed regions. On the other hand, the regions found by intuition are not necessarily IBC. Third, since any state in our proposed safe position-speed regions is reachable from any other state in the safe region through the region itself, in planning a reference trajectory for the robot (a position path and its corresponding scheduled speed), it would then be recommendable to select the states of the reference trajectory inside the proposed regions, which would ensure that they can be reached within the safety constraints and under the actuator limits. On the other hand, one can verify that the state $x_s\in X_I$, shown in Figure \ref{fig:ex_DI_Int}, is not reachable from $0\in X_I^{\circ}$ within $X_I$, i.e. it is not reachable from other states in the safe region through the region itself. Hence, $x_s$ should not be selected as a state in a reference trajectory, as the system cannot reach the state from a safe starting position without violating the constraints. One can see from Figure \ref{fig:ex_DI_lim} that our proposed algorithm automatically excludes these non-reachable parts of the safe region to ensure full controllability on the proposed safe region.

It is noteworthy that all the polytopes in Figures \ref{fig:ex_DI_lim} and \ref{fig:ex_DI_Int} satisfy the controlled invariance property. One can see that the polytope $X_I$ in Figure \ref{fig:ex_DI_Int} forms a larger invariant set compared to the polytopes in Figure \ref{fig:ex_DI_lim}. Indeed, using standard algorithms for calculating the largest polytopic invariant set within the given safety constraints \citep{blanchini}, we end up with a polytope similar to $X_I$ in Figure \ref{fig:ex_DI_Int}. Nevertheless, we emphasize that our proposed algorithm intentionally excludes some parts from the largest controlled invariant set to achieve the advantages mentioned in the previous paragraph.             

\subsection{Ground Robots}
In this subsection, we consider ground robots, modeled as unicycles with acceleration limits. In particular, we have the model
\begin{equation}
\label{eq:uni}
\begin{split}
\dot{x}_1&=x_4cos(x_3)\\
\dot{x}_2&=x_4sin(x_3)\\
\dot{x}_3&=u_2\\
\dot{x}_4&=u_1,
\end{split}
\end{equation}
where $(x_1,x_2)$ is the Cartesian position of the unicycle in a world frame, $x_3$ is the orientation of the unicycle w.r.t. the $x_1$-axis, $x_4$ is the linear driving velocity, $u_1$ is the linear driving acceleration input, and $u_2$ is the steering velocity input. Notice that \eqref{eq:uni} differs from the kinematic model of unicycles, in which it is assumed that one can directly control the linear driving velocity. While it is easy to show that under the kinematic model we can ensure safety of the ground robots since we can decelerate the robot to zero velocity immediately, this is not the case for the more practical model \eqref{eq:uni}. Imagine a scenario in which the robot is initiated at a high linear velocity $x_4$ in the direction of the edges of a given Cartesian region. It may happen that with the limits on the linear acceleration input $u_1$, we cannot decelerate the robot fast enough to avoid collision. Hence, we study here the construction of safe speed profiles for \eqref{eq:uni}. We hereby assume that for low linear velocities, $|x_4|\leq x_{4,min}$, we can safely connect any two states of \eqref{eq:uni} in the given position-velocity limits, and so the problem would be only in operating the robot at high linear velocities.

The system \eqref{eq:uni} can be feedback linearized as follows (Chapter 5 of \citep{Isidori}). By defining the outputs $y_1=x_1$, $y_2=x_2$, and using the feedback linearization law:
\begin{equation}
\label{eq:fb_lin_uni}
%\small
\left[
\begin{array}{c}
u_1  \\
u_2  
\end{array}
\right]=
\left[ 
\begin{array}{cc}
cos(x_3) & -x_4~sin(x_3)  \\
sin(x_3) & x_4~cos(x_3)   
\end{array} 
\right]^{-1} 
\left[ 
\begin{array}{rr}
v_1\\v_2\end{array} 
\right],  
%\normalsize
\end{equation}
we get $\ddot{y}_1=\ddot{x}_1=v_1$ and $\ddot{y}_2=\ddot{x}_2=v_2$, which are two decoupled double integrators representing the dynamics in the two Cartesian directions. Notice that the matrix in \eqref{eq:fb_lin_uni} is invertible at any state except those having $x_4=0$. Thus, for low linear velocities, one should not use \eqref{eq:fb_lin_uni} to avoid the singularity problem. Also, notice that one can define limits on the inputs of the linearized model $v_1,~v_2$ to ensure that the actuator limits of the ground robot, i.e. the limits on $u_1,~u_2$, are satisfied. For instance, suppose that the actuator limits for the ground robot are: $-10\leq u_1\leq 10$ and $-5\leq u_2\leq 5$. Also, suppose that we depend on the feedback linearization law \eqref{eq:fb_lin_uni} in controlling the ground robot as long as the linear velocity $x_4$ is such that $|x_4|\geq \sqrt{2}$ m/s which clearly prevents the discussed singularity problem. With the aid of \eqref{eq:fb_lin_uni}, it can be verified that if $-5\leq v_1\leq 5$ and $-5\leq v_2\leq 5$, then the actuator limits on $u_1,~u_2$ are always satisfied.  

\begin{figure}[t]
\begin{center}
\includegraphics[scale=.32, trim = 10mm 5mm 10mm 5mm]{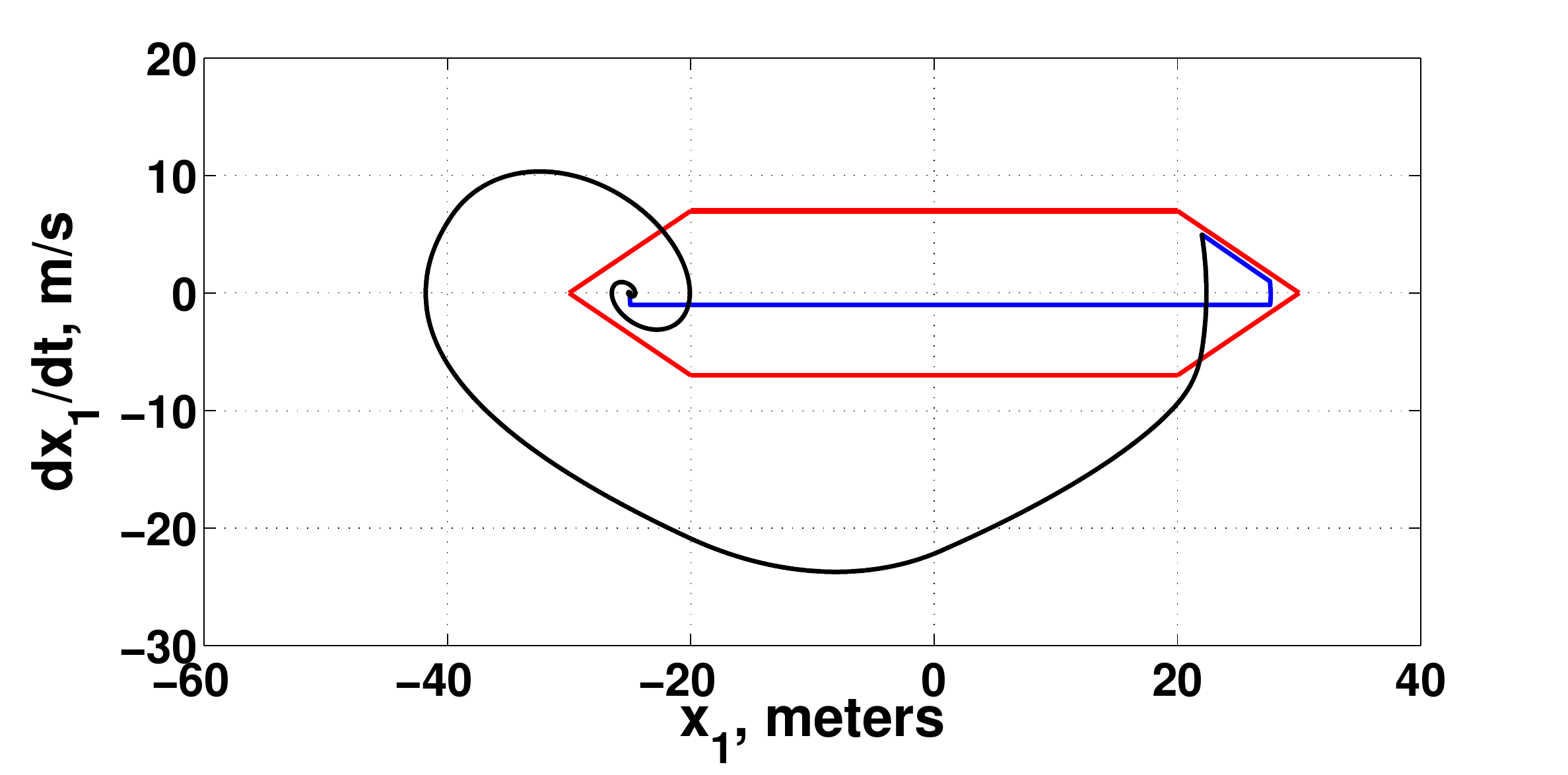} 
\end{center}
\caption{\scriptsize{The proposed safe speed profile in the $x_1$-direction. Blue trajectory: proposed trajectory for achieving constrained mutual accessibility. Black trajectory: standard PD controller, asymptotically stabilizing the desired state.}}
\label{fig1_unicycle}
\end{figure}

\begin{figure}[t]
\begin{center}
\includegraphics[scale=.32, trim = 10mm 5mm 10mm 5mm]{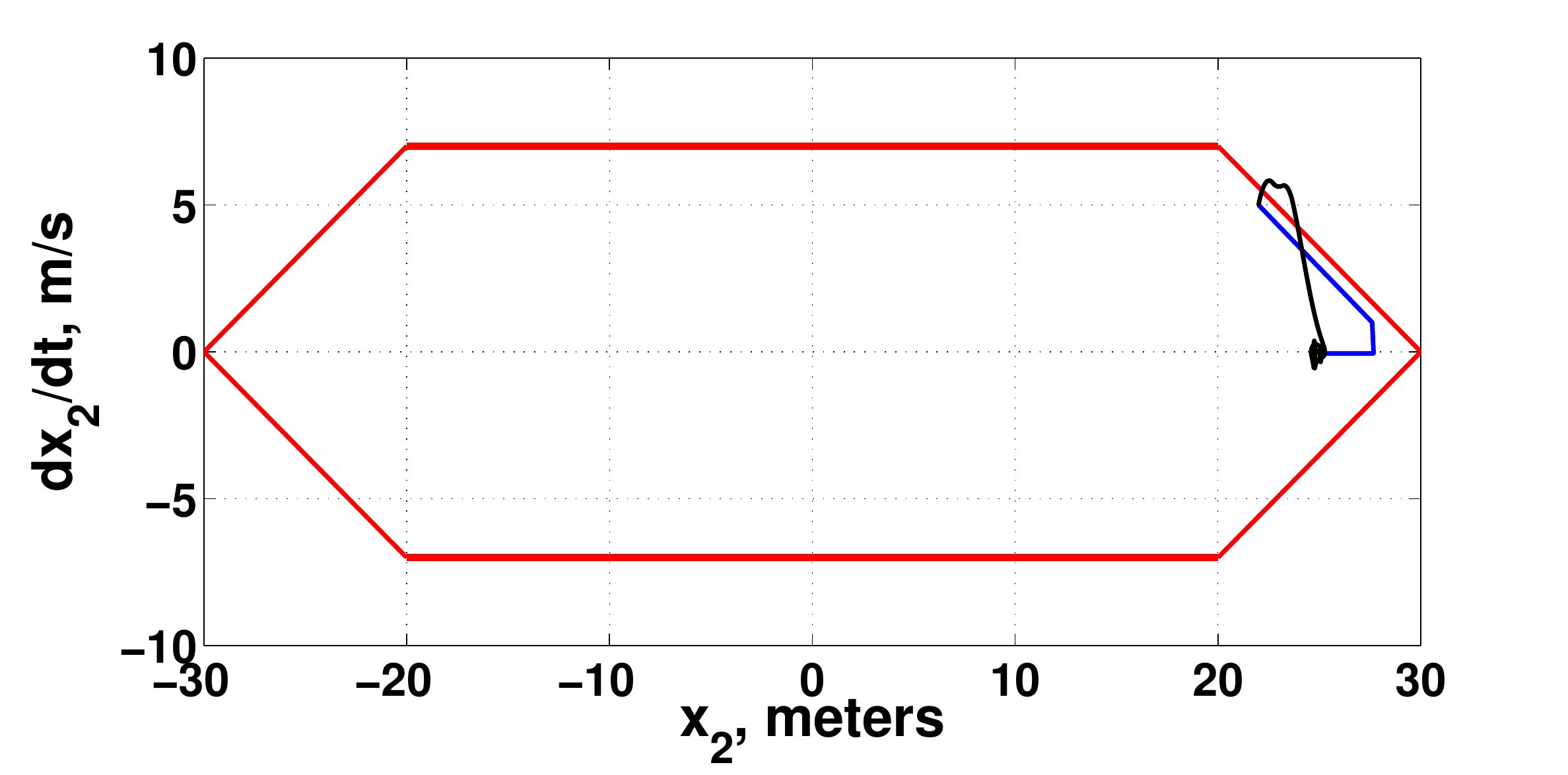} 
\end{center}
\caption{\scriptsize{The proposed safe speed profile in the $x_2$-direction. Blue trajectory: proposed trajectory for achieving constrained mutual accessibility. Black trajectory: standard PD controller, asymptotically stabilizing the desired state.}}
\label{fig2_unicycle}
\end{figure}

Similar to our discussion in the previous subsection, given position and velocity limits in the two Cartesian directions as well as limits on the acceleration inputs $v_1,~v_2$, one can exploit Algorithm \ref{alg_paper1} to construct an IBC region for the linearized system in the new coordinates. The IBC region represents safe speed profiles for the robot in the two Cartesian directions. For instance, suppose that we have the position constraints $-30\leq x_1\leq 30$, $-30\leq x_2\leq 30$, the velocity constraints $-7\leq \dot{x}_1\leq 7$, $-7\leq \dot{x}_2\leq 7$, and the constraints $-5\leq v_1\leq 5$, $-5\leq v_2\leq 5$ obtained from the actuator limits of the ground robot as discussed above. Using Algorithm \ref{alg_paper1}, we construct the safe speed profiles for the two Cartesian directions as shown in red in Figures \ref{fig1_unicycle} and \ref{fig2_unicycle}.     

Now to connect any two states $x_0$ and $x_f$ within the obtained safe region, one can start by finding a connecting trajectory $x(t)$, $t\in [0,t_f]$, for the linearized model. Then, one can depend on the equivalence between the linearized model and \eqref{eq:uni} as long as the linear velocity $x_4$ does not drop to a low value ($|x_4|< \sqrt{2}$ in our example). For the parts of the obtained connecting trajectory $x(t)$ with low linear velocities $x_4$, we avoid using \eqref{eq:fb_lin_uni}, and directly control the nonlinear model \eqref{eq:uni} to connect the two states of the trajectory having low linear velocities, which can always be done safely by assumption as stated at the end of first paragraph in this subsection. Considering again our example, suppose that it is required to connect the state $x_0=(x_1,x_2,\dot{x}_1,\dot{x}_2)=(22,22,5,5)$ to the state $x_f=(x_1,x_2,\dot{x}_1,\dot{x}_2)=(-25,25,0,0)$ in finite time within the given position and velocity constraints and under the actuator limits of the ground robot. Notice that the ground robot is initiated with positive velocities in the directions of the edges $x_1=30$ and $x_2=30$. Figures \ref{fig1_unicycle} and \ref{fig2_unicycle} show, in blue lines, the proposed trajectories that achieve the constrained mutual accessibility. These proposed trajectories are obtained by first applying the PWL feedback, described in the proof of Theorem \ref{thm:main_paper1}, to decelerate the ground robot, and hence avoid violating the safety position and velocity constraints. When $|x_4|< \sqrt{2}$, we stop using \eqref{eq:fb_lin_uni}, and directly control the nonlinear system. In particular, we first decelerate the robot to zero velocity safely, then we set $u_1=0$ and control the input $u_2$ to change the steering angle of the robot, so that the robot heads towards the desired point in the $(x_1,~x_2)$-plane, and finally, we set $u_2=0$ and control $u_1$ to reach the desired point in finite time. It is worth mentioning that there may be more optimal ways to control the robot, but since in this paper we focus on the notion of controllability under constraints and not on optimal controller design, we tried to show the existence of a feasible trajectory achieving the constrained mutual accessibility. Nevertheless, as shown next, standard controllers are usually not sufficient for achieving the constrained mutual accessibility. Figures \ref{fig1_unicycle} and \ref{fig2_unicycle} also show, in black lines, the trajectories initiated at $x_0$ under stabilizing PD controllers, designed for the linearized models in the $x_1$-, $x_2$-directions to stabilize the state $x_f$. One can see that using standard PD controllers, the safety position and speed constraints are not necessarily satisfied in the transient phase, which illustrates the need for the proposed results of the paper.

%Fig_Unicycle_IBC_x2dx2

\subsection{Unmanned Aerial Vehicles (UAVs)}
In this subsection, we utilize our proposed algorithm to construct safe controllable position-speed regions for an important class of UAVs, namely quadrotor vehicles \citep{Angela2}, and then show using experimental results on a Parrot AR.Drone 2.0 platform how these regions can be useful for the safe control of UAVs in confined spaces and under vehicle actuation limits. To that end, we start by reviewing briefly the dynamic model of quadrotor vehicles \citep{Angela2}. The quadrotor vehicle has six degrees of freedom: the translational position $(x,y,z)$, measured in the inertial coordinate frame $O$, and the vehicle Euler angles $(\phi,\theta,\psi)$, rotating the inertial frame into the body-fixed frame $\cV$, where $\phi$ is the roll angle representing (for small angles) the rotation of the quadrotor vehicle around the body $x$-axis, $\theta$ is the pitch angle representing (for small angles) the rotation of the vehicle around the body $y$-axis, and $\psi$ is the yaw angle representing (for small angles) the rotation around the body $z$-axis. Notice that the full state of the quadrotor vehicle also includes the translational velocities $(\dot{x},\dot{y},\dot{z})$ in the inertial frame $O$ and the rotational velocities of the body frame $(p,q,r)$ in $\cV$, and so the dynamic model of the quadrotor vehicle has 12 states. Let $s:=(x,y,z)$. The translational dynamics of the quadrotor vehicle are represented by    
\begin{equation}
\label{eq:UAV1}
\ddot{s} = R_{ZYX}(\psi,\theta,\phi)\bar{f}-\bar{g},
\end{equation}   
where $\bar{f}=(0,0,f)$, $f$ is the sum of the four rotor forces $F_i$ normalized by the vehicle mass $m$, i.e. $f=\sum_{i=1}^{4}(F_i/m)$, $\bar{g}=(0,0,g)$, and $R_{ZYX}(\psi,\theta,\phi)$ is the rotation matrix for transforming coordinates from $\cV$ to $O$, which is given by
\[
R_{ZYX}(\psi,\theta,\phi)=R_z(\psi)R_y(\theta)R_x(\phi),
\]
where 
\[
R_x(\phi)=\left[ 
\begin{array}{ccc}
1 & 0 & 0  \\
0 & cos(\phi) & -sin(\phi) \\
0 & sin(\phi) & cos(\phi)   
\end{array} 
\right],
\]
\[
R_y(\theta)=\left[ 
\begin{array}{ccc}
cos(\theta) & 0 & sin(\theta)  \\
0 & 1 & 0 \\
-sin(\theta) & 0 & cos(\theta)   
\end{array} 
\right],
\]
\[
R_z(\psi)=\left[ 
\begin{array}{ccc}
cos(\psi) & -sin(\psi) & 0  \\
sin(\psi) & cos(\psi) & 0 \\
0 & 0 & 1   
\end{array} 
\right].
\]

The Parrot AR.Drone 2.0 platform has an onboard controller that takes four inputs: the desired pitch angle $\theta_d$, the desired roll angle $\phi_d$, the desired vertical velocity of the vehicle $\dot{z}_d$ and the desired angular velocity of the vehicle around the body's $z$-axis $r_d$, and then it calculates the required four motor forces $F_{i,d}$, $i\in{1,\cdots,4}$. In this paper, we assume that all the states of the quadrotor vehicle are measured. We first use standard, nonlinear controllers to stabilize the $z$-value of the vehicle to a fixed value $z=z_d$, and the yaw angle of the vehicle to zero ($\psi_d=0$). Then, we manipulate $\theta_d$ and $\phi_d$ to control the vehicle's motion in the $x$-, $y$-directions. Assuming that the nonlinear controller successfully stabilizes the vehicle at $z=z_d$ and $\psi=\psi_d=0$, we can assume $\ddot{z}=0$ and $\psi=0$ in the translational dynamics \eqref{eq:UAV1}, and then \eqref{eq:UAV1} can be reduced to
\begin{align}
%\begin{equation}
\label{eq:UAV2}
\ddot{x} &= g\, tan(\theta), \\ 
%\end{equation} 
%\begin{equation}
\label{eq:UAV3}
\ddot{y} &= -g\, tan(\phi)/cos(\theta).
%\end{equation} 
\end{align}
Now we linearize the dynamics \eqref{eq:UAV2} and \eqref{eq:UAV3}, so that we can apply the proposed algorithm in this paper to calculate safe speed profiles for the quadrotor vehicle in the $x$-, $y$-directions. To that end, let $v_1:=g\, tan(\theta_d)$ and $v_2:=-g\, \frac{tan(\phi_d)}{cos(\theta_d)}$. Equivalently, $\theta_d=arctan(\frac{v_1}{g})$ and $\phi_d=arctan(\frac{-v_2cos(\theta_d)}{g})$. If the onboard controller successfully stabilizes the angles $\phi$ and $\theta$ to these selected reference angles $\phi_d$ and $\theta_d$, respectively, then the translational dynamics in the $x$-, $y$-directions become
\begin{align}
%\begin{equation}
\label{eq:UAV4}
\ddot{x} &= v_1, \\
%\end{equation} 
%\begin{equation}
\label{eq:UAV5}
\ddot{y} &= v_2,
%\end{equation}
\end{align}
which are decoupled double integrators. Since the onboard controller typically operates much faster than the position controllers\footnote{In our experiments, the position controllers operate at a $70~$Hz rate, while the onboard controller operates around three times faster.}, it is reasonable to assume that the angles $\phi$ and $\theta$ are stabilized to the desired ones $\phi_d$ and $\theta_d$ quickly, and we can assume that \eqref{eq:UAV4} and \eqref{eq:UAV5} hold approximately. 

Next, we translate the actuator limits on the quadrotor vehicle to constraints on the linearized inputs $v_1,~v_2$. For our quadrotor platform, we have the following constraints on the inputs to the onboard controller: $|\phi_d|\leq 0.32$ rad, and $|\theta_d|\leq 0.32$ rad. It can be verified that if $|v_i|\leq 3.247$, $i\in\{1,2\}$, then the constraints on $\phi_d$ and $\theta_d$ are satisfied. 

\begin{figure}[t]
\begin{center}
\includegraphics[scale=.3, trim = 10mm 5mm 10mm 5mm]{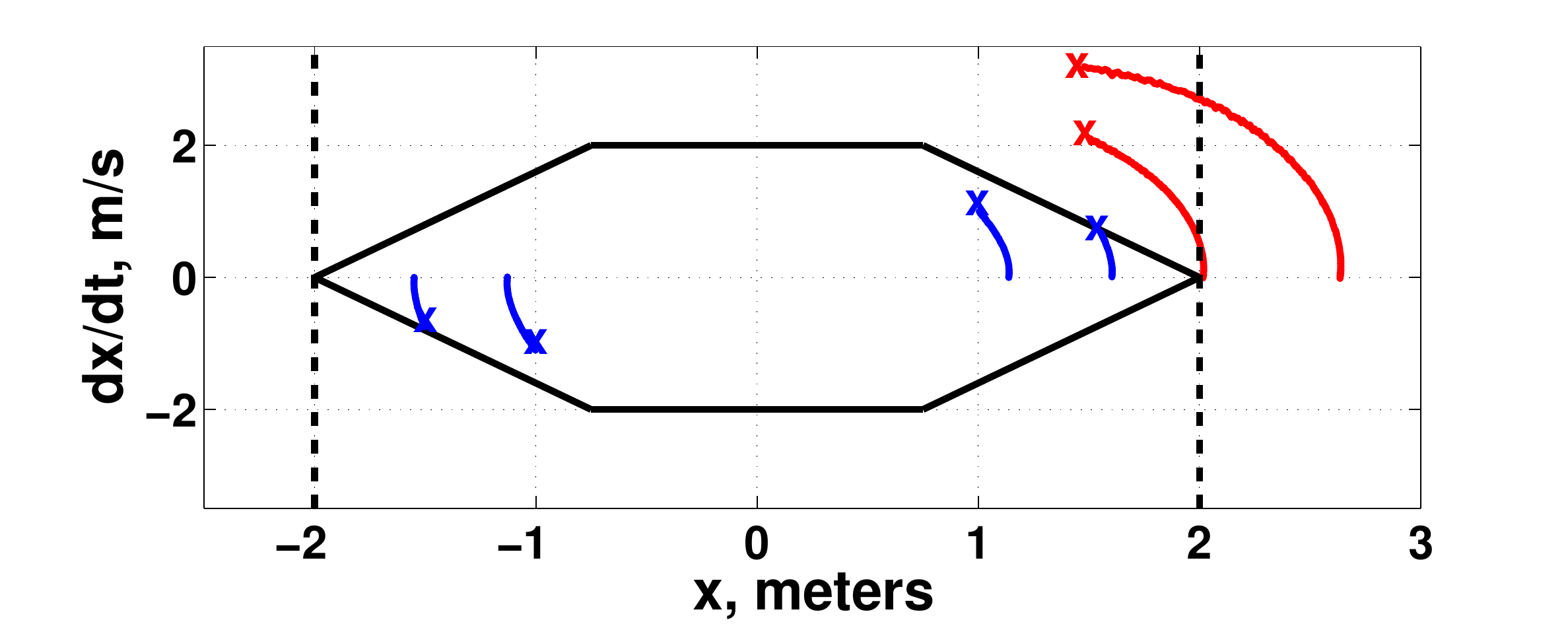} 
\end{center}
\caption{\scriptsize{The IBC region in the $x$-direction for the position safety constraints $-2\leq x\leq 2$ and under the actuation limit $|\theta_d|\leq 0.32$ rad, and samples of actual trajectories starting at different initial states (marked with a cross) inside and outside the IBC region.}}
\label{fig1_UAV}
\end{figure}

Based on the above, our role reduces to constructing for \eqref{eq:UAV4} and \eqref{eq:UAV5} IBC regions under the limits $|v_i|\leq 3.247$, $i\in\{1,2\}$. Suppose, for instance, that the position safety constraints are: $-2\leq x\leq 2$ and $-2\leq y \leq 2$. Similar to Example \ref{ex_alg} and the previous subsections, we use Algorithm \ref{alg_paper1} to construct the IBC polytopes. Figure \ref{fig1_UAV} shows the IBC region for the dynamics in the $x$-direction under the limit  $|v_1|\leq 3.247$. Similarly, one can construct an IBC region for the dynamics in the $y$-direction under $|v_2|\leq 3.247$. As discussed before, the IBC region in Figure \ref{fig1_UAV} provides for each position a corresponding safe speed range. If one limits the speed at any position $x$, $-2\leq x \leq 2$, to the safe speed range, then there exist feasible control inputs that keep the state trajectory inside the IBC region, and prevent the violation of the position safety constraints. Moreover, we provide in the proof of Theorem \ref{thm:main_paper1} of the paper a constructive method for synthesizing a PWL feedback that keeps the state trajectories inside the IBC regions. 

In our first experiment, we stabilize the $y$-value of the quadrotor vehicle to $0$, the $z$-value to $z_{d}=1.5$, and the yaw angle to $\psi_d=0$. We also allow the vehicle to gain an initial velocity in the direction of the edge of the position range $-2\leq x \leq 2$, and then initiate our proposed PWL feedback in the proof of Theorem \ref{thm:main_paper1} to decelerate the quadrotor vehicle to zero velocity, and so prevent the vehicle from violating the position constraints. Figure \ref{fig1_UAV} shows samples of the state trajectories, under the proposed feedback, initiated at different critical states inside the IBC region (blue trajectories). One can see that for all the shown, critical initial conditions in the IBC region, the proposed feedback successfully keeps the state trajectories in the IBC region, and prevents the violation of the safety constraints. After decelerating the vehicle to zero velocity, one can apply a robust hovering controller to keep the vehicle in place, or one can safely drive the vehicle to a safe point within the position constraints (e.g., the center of the safe region). Figure \ref{fig1_UAV} also shows two cases where the vehicle is initiated at high initial velocities, outside the safe speed profile, in the direction of the edge $x=2$, but with initial positions inside of the position constraints (red trajectories). One can see that for these cases the proposed feedback, built based on the vehicle's actuator limits, cannot decelerate the vehicle fast enough, and the position safety constraints are violated. The experiment shows the importance of keeping the quadrotor vehicle's speed within the safe speed profile for preventing collisions.

\begin{figure}[t]
\begin{center}
\includegraphics[scale=.32, trim = 10mm 5mm 10mm 10mm]{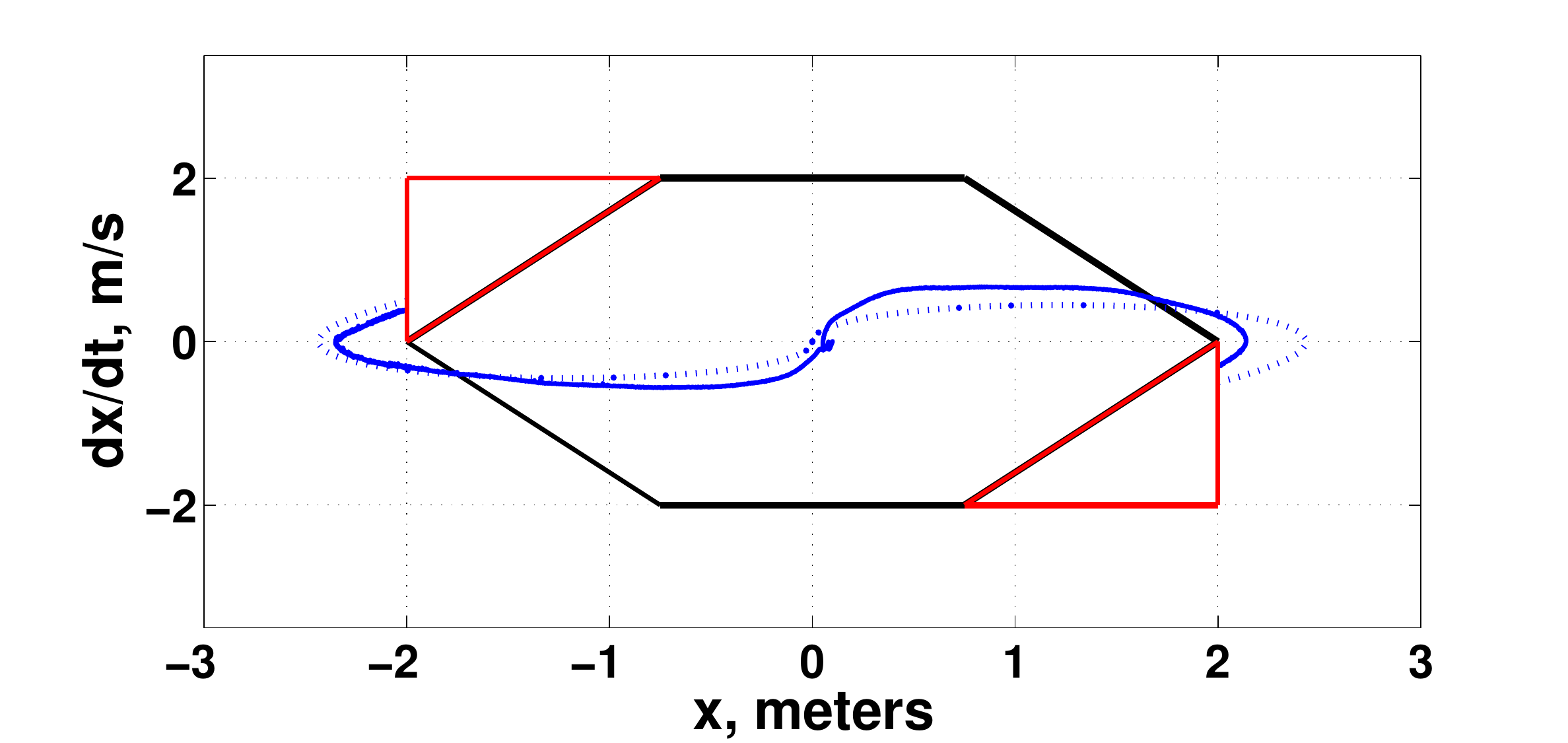} 
\end{center}
\caption{\scriptsize{The state trajectories connecting the origin to some points outside the IBC region (dotted lines: simulations; solid lines: experiments). Points within the red triangles are not reachable from other points in the safe region through the region itself.}}
\label{fig2_UAV}
\end{figure}

In the second experiment, we compare the proposed safe speed profile to the ones that can be obtained by intuition or by the controlled invariance property. One can argue that the states in the red triangles in Figure \ref{fig2_UAV} should be included in the safe position-speed region since starting from any state in the red triangles, the position constraints are not violated. For instance, starting very close to the edge $x=2$ with a high negative velocity will make the vehicle head towards the other edge, and the position constraints are not violated. However, these states in the red triangles are not reachable from all other states inside the safe region through the region itself. Hence, our algorithm automatically truncates these red triangles to ensure full controllability on the safe region. In Figure \ref{fig2_UAV}, we show the state trajectories of connecting the origin to some points in the red triangles. The dotted blue trajectories are obtained from simulation by applying the standard open-loop control law of connecting two states based on the control Gramian (equation (15) of \citep{HC14}, with $t_f=10~s$).  
The solid blue trajectories are real, experimental trajectories obtained by applying similar acceleration profiles to the real system. One can see that the quadrotor vehicle cannot reach the points in the red triangles without violating the safety position constraints. Hence, it is always recommended that the points of reference trajectories are selected inside the IBC region to ensure that they can be reached from other safe states with trajectories that completely lie inside the safe region. 

\begin{figure}[t]
\begin{center}
\includegraphics[scale=.32, trim = 10mm 5mm 10mm 10mm]{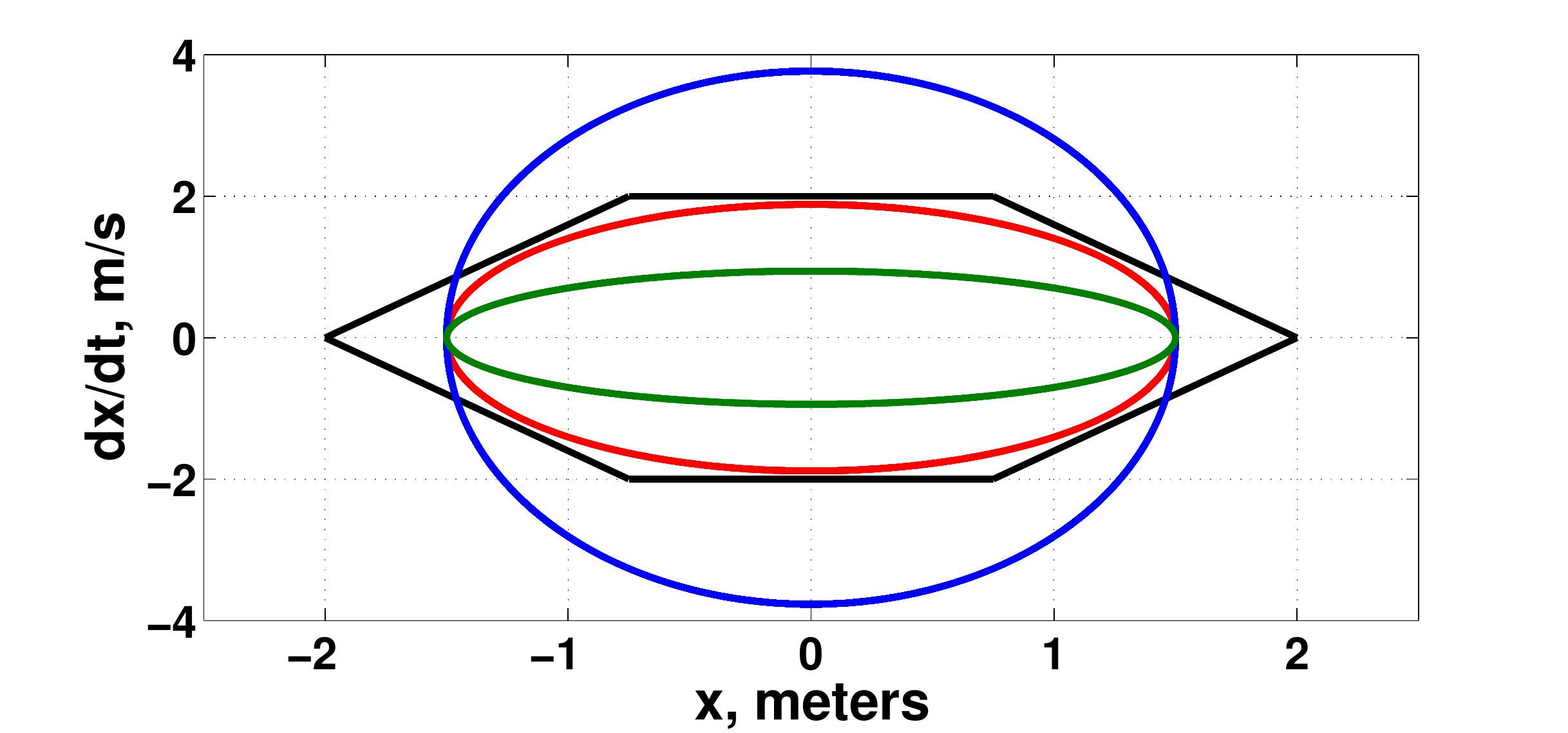} 
\end{center}
\caption{\scriptsize{The speed profile of the reference signal in the $x$-direction for three different frequencies: $0.1\,$Hz (green), $0.2\,$Hz (red), $0.4\,$Hz (blue). The IBC region is shown in black.}}
\label{fig3_UAV}
\end{figure}

\begin{figure}[t]
\begin{center}
\includegraphics[scale=.33, trim = 10mm 5mm 10mm 10mm]{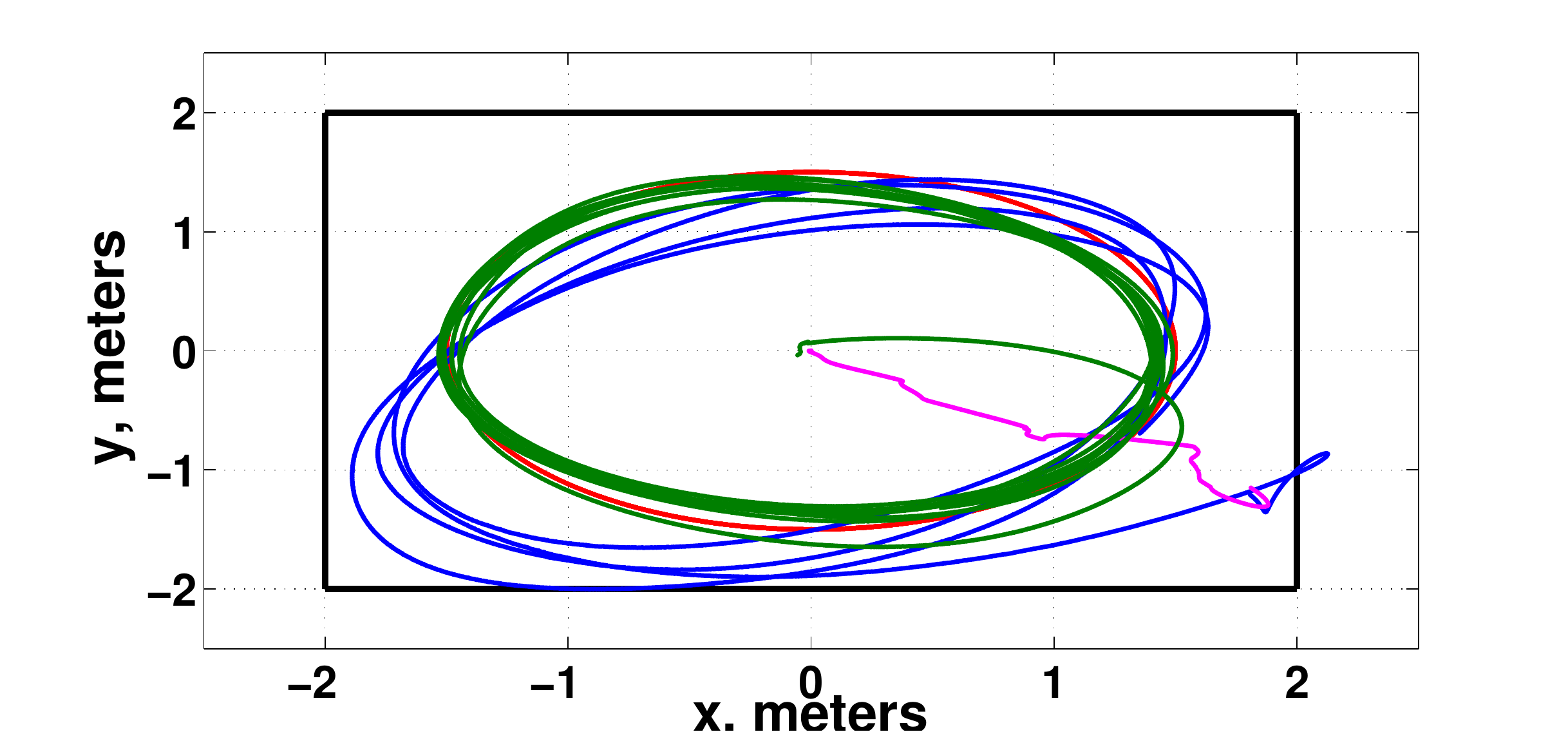} 
\end{center}
\caption{\scriptsize{The tracking of the circle in the $(x,~y)$-plane under standard tracking controllers and starting from two initial conditions. Red line: The reference trajectory in the $(x,~y)$-plane (the circle). Green line: starting from the origin. Blue line: starting from the initial state $(x,y,\dot{x},\dot{y})=(1.8,-1.2,0.31,-0.46)$. Magenta line: the proposed trajectory connecting the initial state to the origin before starting the standard tracking controller.}}
\label{fig4_UAV}
\end{figure}

Thirdly, we show how the constructed IBC regions can be useful in determining feasible reference trajectories for the quadrotor vehicle given the position safety constraints and the vehicle's actuator limits. For example, suppose that it is required to track a circle in the $(x,~y)$-plane of radius 1.5 m, centered at $(0,0)$. This can be achieved by tracking sinusoidal signals in both $x$-, $y$-directions, with a $90^{\circ}$ phase shift. Also, suppose that we have three possible frequencies for going through the circle: $0.1~$Hz, $0.2~$Hz, or $0.4~$Hz. Using the constructed safe speed profiles in the  $x$-, $y$-directions, represented by the IBC regions, we want to determine a suitable, safe frequency. Figure \ref{fig3_UAV} shows, for instance, the speed profile of the reference sinusoidal signal in the $x$-direction for the three frequencies. A similar figure can be drawn for the $y$-direction. One can see that for the $0.4~$Hz rate, the reference speed for the quadrotor vehicle does not completely lie within the safe speed profile, and so we avoid this frequency. Also, the reference speed for the $0.2~$Hz rate is within the safe speed profile and it is faster than the $0.1~$Hz rate. Thus, we select the $0.2~$Hz rate as reference trajectory. Figure \ref{fig4_UAV} shows the tracking of the reference signal in the $(x,~y)$-plane under standard tracking controllers and starting from two initial conditions. The trajectory starting at $(0,0)$, shown in green, remains within the safety position constraints as required. Notice that achieving perfect asymptotic tracking of reference signals is out of the scope of this paper. It is worth mentioning that although we know from the IBC property that there exist uniformly bounded control inputs connecting the initial state $(x,y,\dot{x},\dot{y})=(1.8,-1.2,0.31,-0.46)$ to the states of the reference trajectory within the IBC region, standard tracking controllers may not achieve asymptotic tracking with trajectories that completely lie within the safe position constraints (see, for instance, the blue line in Figure \ref{fig4_UAV}). Hence, for the cases where the quadrotor vehicle is initiated at risky initial conditions (with a positive velocity in the direction of the position region edge), it is recommended that one first applies the PWL feedback in the proof of Theorem~\ref{thm:main_paper1} to safely decelerate the vehicle, and then connects the vehicle to the point $(0,0)$ in the center of the position region, and finally, applies the standard tracking controller (see the magenta line representing the real trajectory connecting the initial position to the origin within the safe position region). There may be more advanced controller designs that achieve the tracking objective, but as mentioned before, we focus in this paper on controllability under constraints and use the experiments to illustrate the concept.  

\begin{figure}[t]
\begin{center}
\includegraphics[scale=.32, trim = 10mm 5mm 10mm 10mm]{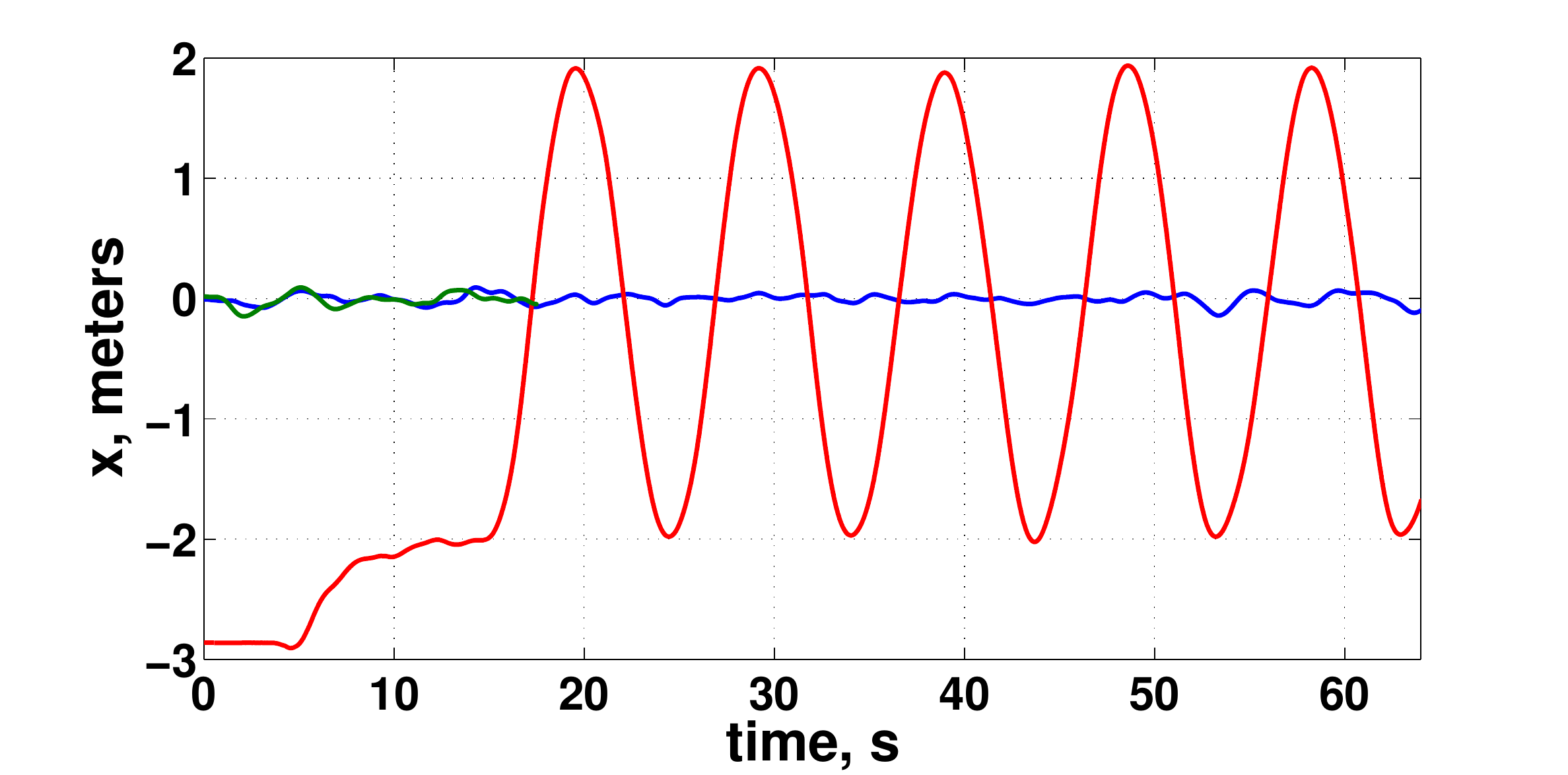} 
\end{center}
\caption{\scriptsize{The $x$-values of the two vehicles. Blue line: Our controlled quadrotor vehicle with the proposed algorithm for updating the safe speed profile. Green line: Our controlled quadrotor vehicle without the proposed algorithm. Red line: The other quadrotor vehicle (the dynamic obstacle).}}
\label{fig5_UAV}
\end{figure}

\begin{figure}[t]
\begin{center}
\includegraphics[scale=.32, trim = 10mm 5mm 10mm 10mm]{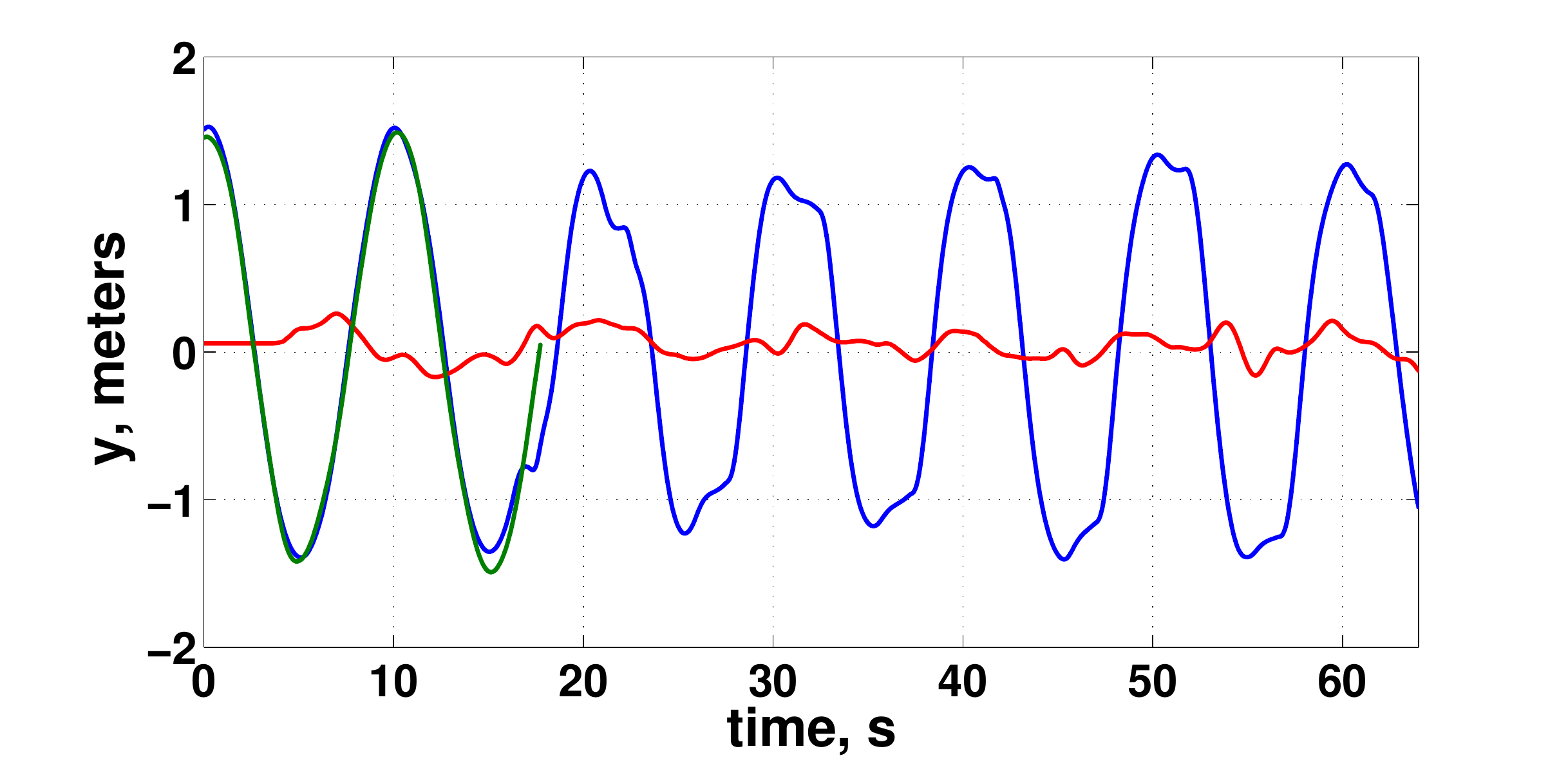} 
\end{center}
\caption{\scriptsize{The $y$-values of the two vehicles. Blue line: Our controlled quadrotor vehicle with the proposed algorithm for updating the safe speed profile. Green line: Our controlled quadrotor vehicle without the proposed algorithm. Red line: The other quadrotor vehicle (the dynamic obstacle).}}
\label{fig6_UAV}
\end{figure}

\begin{figure}[t]
\begin{center}
\includegraphics[scale=.32, trim = 10mm 5mm 10mm 10mm]{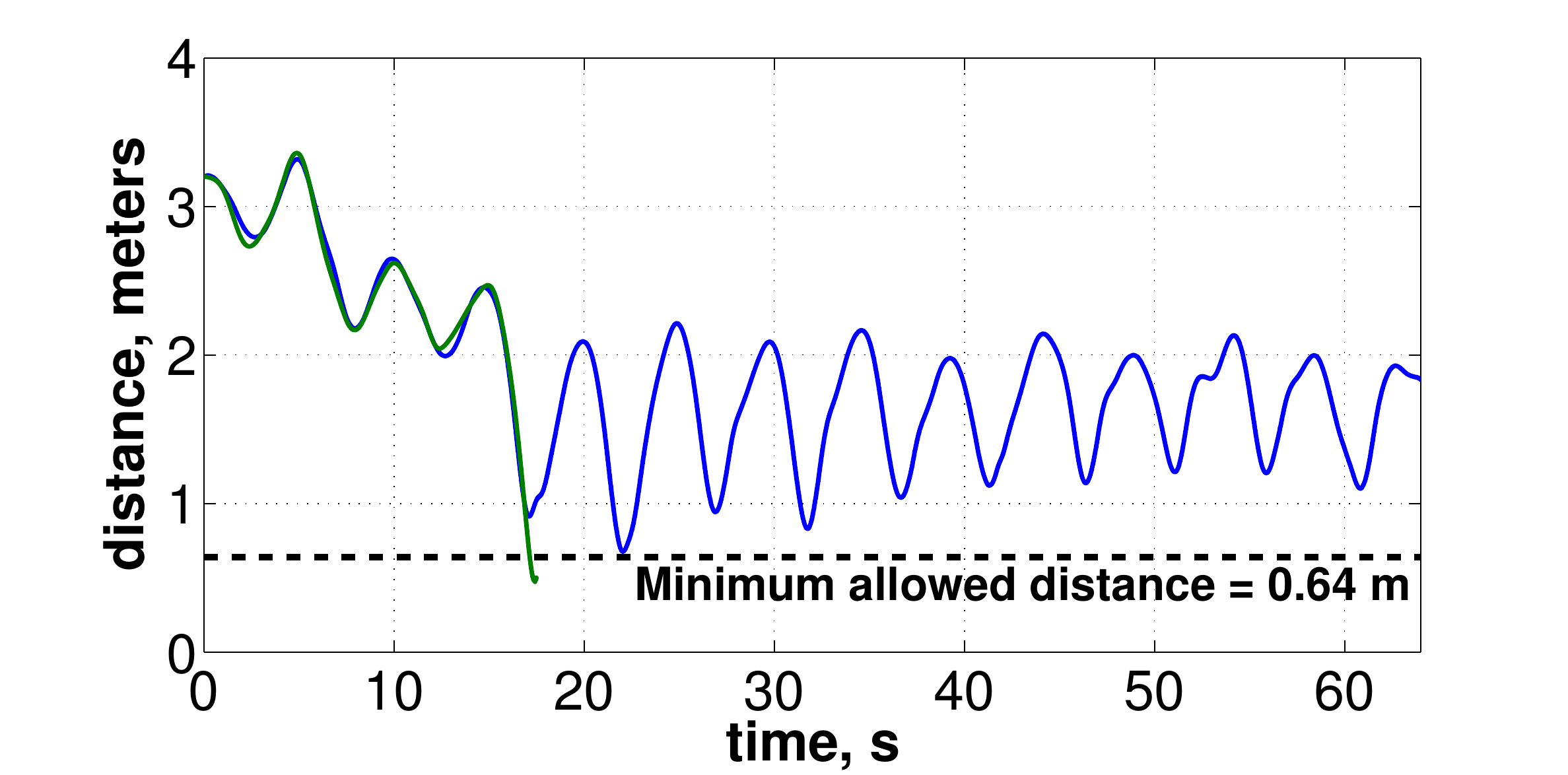} 
\end{center}
\caption{\scriptsize{The Euclidean distance between the $x$-, $y$-coordinates of the two vehicles ($\sqrt{(x_1-x_2)^2+(y_1-y_2)^2}$). Blue line: with the proposed algorithm for updating the safe speed profile online. Green line: without updating the safe speed profile online (the experiment was stopped after collision).}}
\label{fig7_UAV}
\end{figure}

As discussed before, one advantage of the proposed algorithm for constructing the IBC regions is that it is computationally efficient. This enables us to implement the algorithm in real time to update the safe speed profile online if the position constraints change. To illustrate the main idea, in the next set of experiments, we use the proposed algorithm to achieve dynamic obstacle avoidance when the dynamic obstacles intersect with the quadrotor vehicle's path. The idea is to detect that a dynamic obstacle is about to intersect with the quadrotor vehicle's path, and then to update the position constraints accordingly. Then, our proposed algorithm is applied in real time to calculate for the updated position constraints a corresponding safe speed profile\footnote{The safe speed profile is updated at each sampling instant, where the sampling frequency is $70~$Hz.}, as well as an associated PWL feedback, as proposed in the proof of Theorem \ref{thm:main_paper1}, which keeps the quadrotor vehicle within the updated safety position constraints. In the fourth experiment, we let the quadrotor vehicle track a sinusoidal reference trajectory in the $y$-direction with a frequency of $0.1~$Hz, while stabilizing the $x$-value to $x_d=0$ and keeping a constant height. We then run another quadrotor vehicle, our dynamic obstacle, to track a sinusoidal reference trajectory in the $x$-direction with a frequency of $0.1~$Hz, while stabilizing the $y$-value to $y_d=0$ and again keeping a constant height. The two vehicles collide if their $x$-, $y$-coordinates coincide. 
%, i.e one vehicle would lie below the other one in the $z$-direction. Indeed, this would cause a downwash force on the lower vehicle, and may cause its collision with the ground. Hence, we want to avoid this scenario. 
To achieve collision avoidance, we run the proposed algorithm for the first quadrotor vehicle in real time to update the safe speed profile online. Figures \ref{fig5_UAV}, \ref{fig6_UAV} and \ref{fig7_UAV}  show that the proposed algorithm succeeds and prevents the collision between the two vehicles\footnote{The radius of the quadrotor body is $0.32~$m, and so the Euclidean distance between the $x$-, $y$-coordinates of the two vehicles ($\sqrt{(x_1-x_2)^2+(y_1-y_2)^2}$) should be kept higher than $0.64~$m.}. To emphasize the effect of our proposed algorithm, we also illustrate in Figures \ref{fig5_UAV}, \ref{fig6_UAV} and \ref{fig7_UAV} in green 
%lines the change in the $x,~y$ values of the first vehicle and the distance between the two vehicles for 
the case where we do not run our proposed algorithm to update the safe speed profile online. One can see that the Euclidean distance between the $x$-, $y$-coordinates of the two vehicles drops below $0.5~$m in this case, which we consider a crash given the vehicle body radius of $0.32~$m.
%, i.e. one vehicle almost lies below the other one which we consider as a collision.

\begin{figure}[t]
\begin{center}
\includegraphics[scale=.32, trim = 10mm 5mm 10mm 10mm]{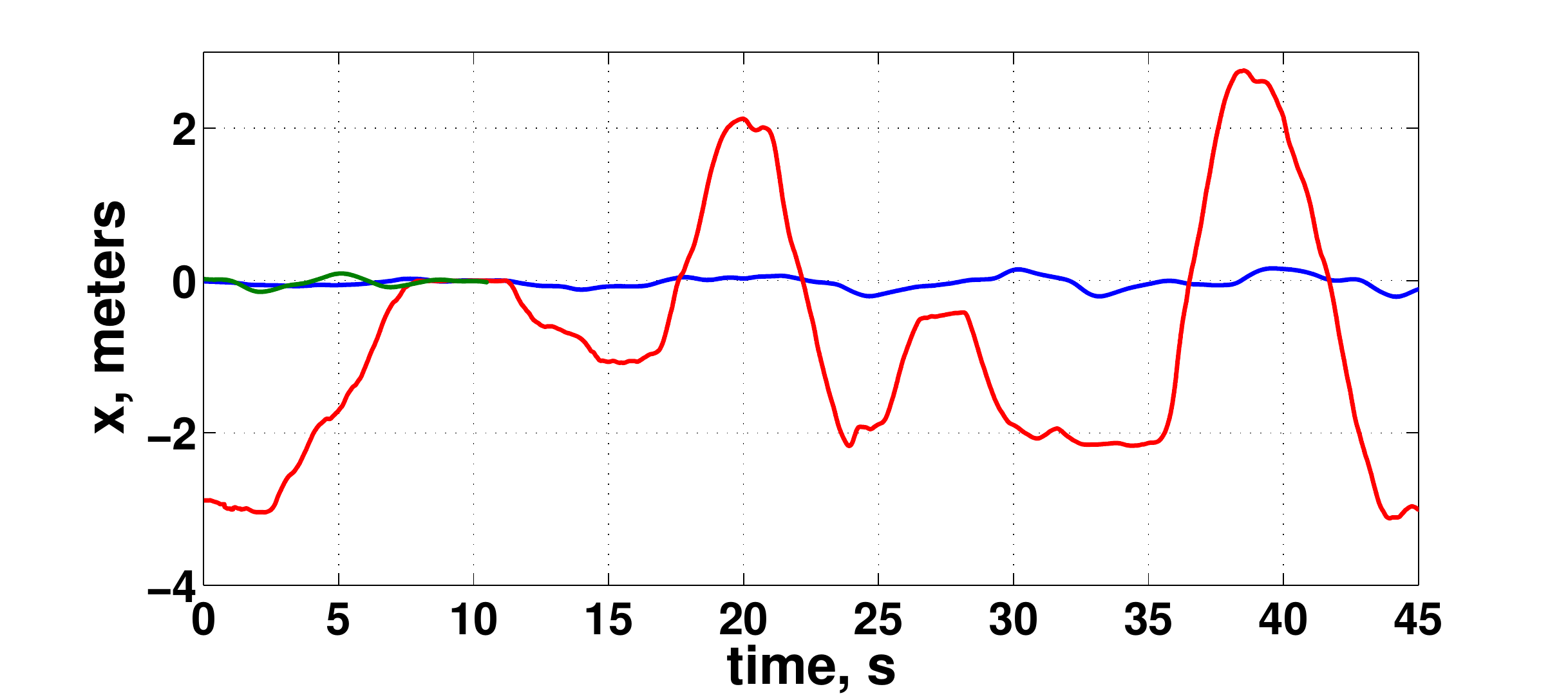} 
\end{center}
\caption{\scriptsize{The $x$-values of the quadrotor vehicle and the moving human. Blue line: quadrotor vehicle with the proposed algorithm of updating the safe speed profile online. Red line: human motion. Green line: quadrotor vehicle without the proposed algorithm.}}
\label{fig8_UAV}
\end{figure}

\begin{figure}[t]
\begin{center}
\includegraphics[scale=.32, trim = 10mm 5mm 10mm 8mm]{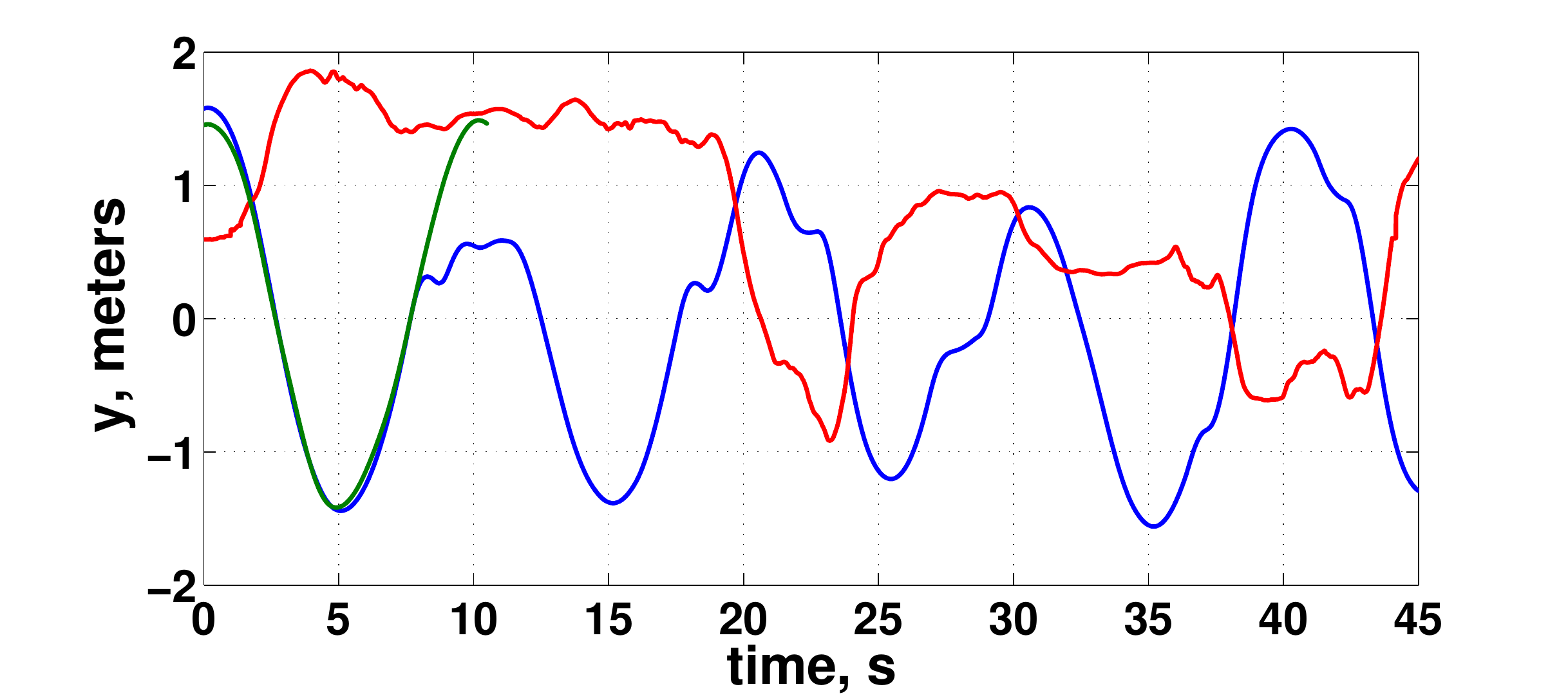} 
\end{center}
\caption{\scriptsize{The $y$-values of the quadrotor vehicle and the moving human. Blue line: quadrotor vehicle with the proposed algorithm of updating the safe speed profile online. Red line: human motion. Green line: quadrotor vehicle without the proposed algorithm.}}
\label{fig9_UAV}
\end{figure}

\begin{figure}[t]
\begin{center}
\includegraphics[scale=.32, trim = 10mm 5mm 10mm 8mm]{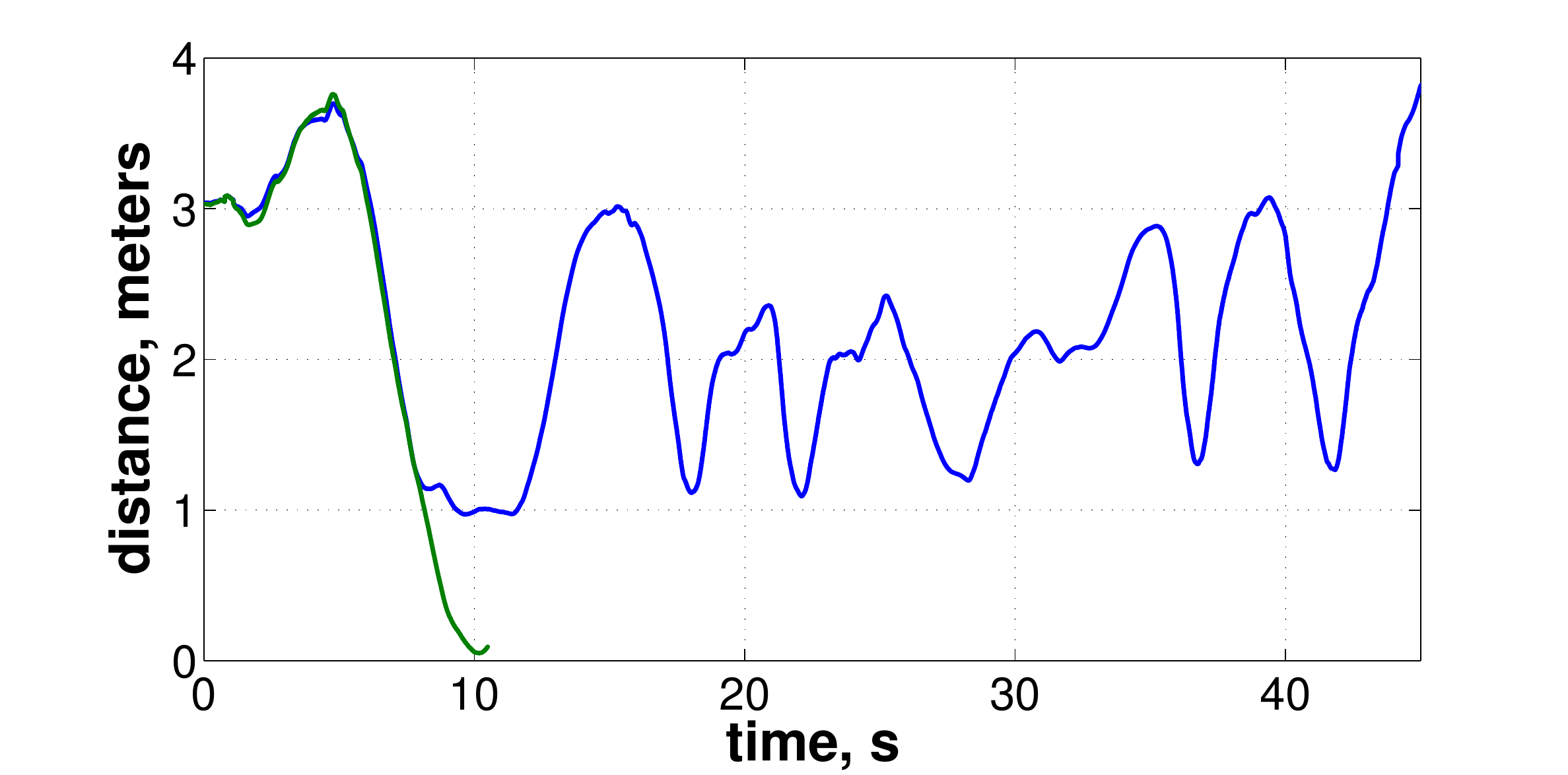} 
\end{center}
\caption{\scriptsize{The Euclidean distance between the $x$-, $y$-coordinates of the quadrotor vehicle and the moving human. Blue line: with the proposed algorithm for updating the safe speed profile online. Green line: without updating the safe speed profile online (the experiment was stopped after collision).}}
\label{fig10_UAV}
\end{figure}

In Figures \ref{fig8_UAV}, \ref{fig9_UAV} and \ref{fig10_UAV}, we repeat the same experiment after replacing the second quadrotor vehicle with a random human motion that intersects with the path of our controlled quadrotor vehicle. One can see from the figures that our proposed algorithm succeeds to prevent the collision with the moving human\footnote{A demo video can be found at:\\ \url{https://drive.google.com/folderview?id=0BxQ2msoW3w5sampMRG9rZzluVjg&usp=sharing}.}. 

\section{Conclusions}

\label{sec:con}

In this paper, we studied the problem of constructing IBC regions for affine systems. That is, we construct safe regions in the state space within which we can fully control the given affine system using uniformly bounded control inputs. After formulating the problem, we discussed the difficulties that are faced if one tries to directly exploit the existing results for checking IBC on given polytopic regions. Instead, we explored the geometry of the problem, provided a computationally efficient algorithm for constructing IBC regions, and proved its soundness. As sample case studies, we showed how our proposed algorithm can be useful for constructing safe speed profiles for different classes of robotic systems. We also provided several experimental results to verify the theoretical contributions of the paper. This includes using our proposed algorithm for real-time collision avoidance for UAVs.

%In this paper, we have initiated the study of the problem of constructing in-block controllable covers of nonlinear systems on polytopes, which is useful in studying approximate mutual accessibility problems of nonlinear systems on polytopes \citep{HC15}. After formulating the problem, we have discussed the difficulties that may appear if one tries to directly apply the main result of \citep{HC14} to construct the in-block controllable polytopic covers. Then we provided geometric guidelines that help with building such covers, without the need for using trial and error. Moreover, for a special class of nonlinear systems with $n-1$ inputs, where $n$ is the dimension of the system, we have provided a constructive algorithm for building a hierarchy of in-block controllable polytopic covers, and we proved its soundness. We have also provided illustrative examples to clarify the main results of the paper. An open research problem is to extend the algorithm of this paper to more general classes of nonlinear systems.  

%Several illustrative examples have been provided to clarify the main results of the paper. 
%\begin{comment}
\section*{Appendix}
\begin{figure}[t]
\begin{center}
\includegraphics[scale=.23, trim = 10mm 152mm 10mm 70mm]{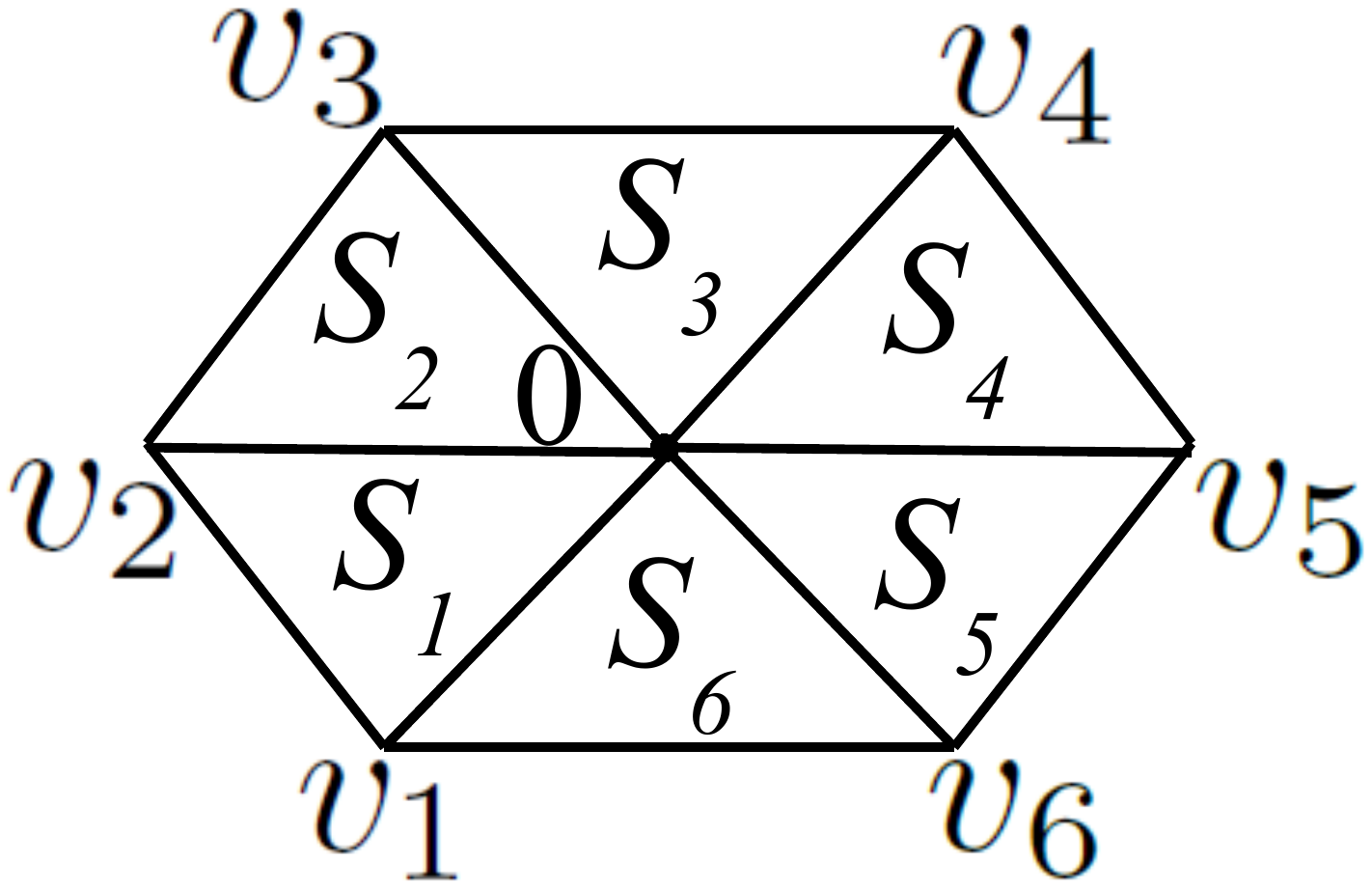} 
\end{center}
\caption{\scriptsize{An illustrative figure for the triangulation used in the proof of Theorem \ref{thm:main_paper1}.}}
\label{fig:triang}
\end{figure}

\textbf{Continuation of the Proof of Theorem \ref{thm:main_paper1}:}
We construct a continuous PWL feedback $u_p(x)$ under which all the trajectories initiated in $X^{\circ}$ tend to $\cO$ through $X^{\circ}$. At a vertex $\bar{v}\in \cO$, select input $\bar{u}$ such that $A\bar{v}+B\bar{u}=0$, which is always possible by the definition of $\cO$. Next, for the vertices $v_i$ satisfying $\cB\cap C^{\circ}(v_i)\neq \emptyset$, identify $\bar{b}_i\in \cB \cap C^{\circ}(v_i)$. Since $\bar{b}_i\in C^{\circ}(v_i)$, then by definition $h_j\cdot \bar{b}_i<0,$ for all $j\in J(v_i)$. Also, since $\bar{b}_i\in \cB$, there exists $\bar{u}_i \in \RR^m$ such that $B\bar{u}_i=\bar{b}_i$. Now for $u_i=c_i\bar{u}_i \in \RR^m$, where $c_i>0$, we have
\begin{equation}
\label{eq:lem3_pf}
h_j\cdot (Av_i+Bu_i)=h_j\cdot Av_i + c_i h_j\cdot \bar{b}_i, 
\end{equation}
for all $j\in J(v_i)$. The second term of the right-hand side of \eqref{eq:lem3_pf} is always negative, and we can always select $c_i>0$ sufficiently large such that $h_j\cdot (Av_i+Bu_i)<0$, for all $j\in J(v_i)$. The above control assignment at the vertices of $X$ satisfies the invariance conditions, and for the vertices having $v_i\notin \cO$ and $\cB\cap C^{\circ}(v_i)\neq \emptyset$, it satisfies the invariance conditions strictly (with strict inequalities). At $x=0$, select $u=0$.  
We construct a special triangulation of $X$ using the point set $\{v_1,\cdots,v_p,0\}$, where $\{v_1,\cdots,v_p\}$ are the vertices of $X$, such that if $S_i$ is an $n$-dimensional simplex in the triangulation, then $0\in S_i$ is a vertex of $S_i$ \citep{LEE}. This can be carried out by triangulating each facet of $X$, $F_j$, into $(n-1)$-dimensional simplices, and then taking the convex hull of $0\in X^{\circ}$ and the $(n-1)$-dimensional simplices to form a triangulation of $X$ consisting of $n$-dimensional simplices $S_i$ with the desired property. Figure \ref{fig:triang} shows a $2D$ illustration of the triangulation.  
Based on the control values selected at $\{v_1,\cdots,v_p,0\}$, one can always construct on each simplex $S_i$ a unique affine feedback $k_ix+g_i$. Moreover, $[k_i~g_i]^T=[\bar{V}~\bar{1}]^{-1}\bar{w}$, where $\bar{V}$ is a matrix whose rows are the transpose of the vertices of $\cS_i$, $\bar{1}$ is a column of ones, and $\bar{w}$ is a column of the transpose of the selected inputs at the vertices of $\cS_i$ \citep{HVS04}. Since $u=0$ at $x=0$ by assignment and $0\in \cS_i$, then $g_i=0$; that is, the feedback on each $\cS_i$ is linear. It can be easily shown that the overall control law is a continuous PWL feedback, denoted by $u_p(x)$, and by a simple convexity argument, $u_p(x)$ satisfies the invariance conditions of $X$ at every $x\in \partial X$ \citep{HVS04}.%, the boundary of $X$ \citep{HVS04}. 

Let $f(x):=Ax+Bu_p(x)$. Since $0$ is a vertex in each $S_i$, $f(0)=0$ and $f(x)$ is linear on each $S_i$, then the vector field on $\partial(\lambda X)$ represents $\lambda$-scaled vectors of the vector field on $\partial X$ for any $\lambda\in (0,1)$. Therefore, $u_p(x)$ satisfies the invariance conditions of $\lambda X$ for any $\lambda \in (0,1)$, and so starting from any $x_0\in \lambda X$, $\phi(x_0,t,u_p)\in \lambda X$ for all $t\geq 0$.

We next show that for every $x_0\in X^{\circ}$, $\phi(x_0,t,u_p) \rightarrow \cO$ as $t\rightarrow \infty$, which implies by a simple argument that we can steer the trajectories to an $\epsilon$-neighborhood of $\cO$ in finite time, where $\epsilon>0$ can be selected arbitrarily small.
Since $0\in X^{\circ}$ by assumption, it is known that $X$ can be expressed as $X=\left\{x\in \RR^n~:~n_i\cdot x\leq 1,~i=1,\cdots,r\right\}$, where $n_i\in \RR^n$, $n_i\cdot x=1$ if $x\in F_i$ and $n_i\cdot x<1$ if $x\in X,~x\notin F_i$. We define $V(x)=\max_{i\in\left\{1,\cdots,r\right\}}n_i\cdot x$. Notice that if $x\in \partial X$, then $V(x)=1$. Similarly, if $x\in \partial(\lambda X)$ for $\lambda\in (0,1)$, then $V(x)=\lambda$. One can show that $V(x)$ is locally Lipschitz, and its upper right Dini derivative $D^+_fV(x)=\max_{i\in I(x)}n_i\cdot (Ax+Bu_p(x))$, where $I(x)=\left\{i\in\left\{1,\cdots,r\right\}:n_i\cdot x=V(x)\right\}$ \citep{Danskin}. With the aid of invariance conditions, it is shown in Lemma 5.3 of \citep{HC14}, which also applies to non-simplicial polytopes, that $D^+_fV(x)\leq 0$ for each $x\in X$. We hereby show that additionally $\{x\in X~:~D^+_fV(x)= 0\}\subset \cO$. Notice that for a vertex $v_i\in \cO$, $f(v_i)=0$ by assignment, and so $D^+_fV(v_i)= 0$. Next, the rest of the vertices of $X$ satisfy $\cB\cap C^{\circ}(v_i)\neq \emptyset$ by assumption, and we assigned the control inputs at these vertices to satisfy the invariance conditions strictly. Thus, $n_j\cdot (Av_i+Bu_p(v_i))<0$, for all $j\in J(v_i)$. Note that $j\in I(v_i)$ if by definition $n_j\cdot v_i = V(v_i)=1$, i.e. $v_i\in F_j$. Then by the strict invariance conditions, we have $n_j\cdot (Av_i+Bu_p(v_i))<0$ for all $j\in I(v_i)$, and so $D^+_fV(v_i)<0$ for all the vertices $v_i\notin \cO$. Let $\bar{x}\in \partial X$ be arbitrary, and suppose that $\bar{x}\in S_k$. Let $S_{\bar{x}}$ denote the smallest sub-simplex of $S_k$ such that $\bar{x}\in S_{\bar{x}}^{\circ}$, the relative interior of $S_{\bar{x}}$. Since $\bar{x}\in S_{\bar{x}}^{\circ}$, we can write $\bar{x}=\sum_{s}\alpha_{s}v_s$, where $\alpha_s>0$, $\sum_{s}\alpha_s=1$, and $v_s$ are the vertices of $S_{\bar{x}}$, which are a subset of the vertices of the $n$-dimensional simplex $S_k$. Since the vector field $f(x)$ is linear on the simplex $S_k$ by construction, we have $f(\bar{x})=\sum_{s}\alpha_{s}f(v_s)$. We now study $D^+_fV(\bar{x})$. It is straightforward to show $I(\bar{x})\subset I(v_s)$ for every vertex $v_s\in S_{\bar{x}}$. Then,
\begin{equation*}
\begin{split}
%D^+_fV(\bar{x})=\max_{i\in I(\bar{x})}n_i\cdot f(\bar{x})=\max_{i\in I(\bar{x})}n_i\cdot \sum_{s}\alpha_{s}f(v_s)\\ 
%\leq \sum_{s}\alpha_{s} \max_{i\in I(\bar{x})}n_i\cdot f(v_s)\leq \sum_{s}\alpha_{s} \max_{i\in I(v_s)}n_i\cdot f(v_s)\\ =\sum_{s}\alpha_{s}D^+_fV(v_s).  
D^+_fV(\bar{x})&=\max_{i\in I(\bar{x})}n_i\cdot \sum_{s}\alpha_{s}f(v_s) 
\\&\leq \sum_{s}\alpha_{s} \max_{i\in I(\bar{x})}n_i\cdot f(v_s)\\ &\leq \sum_{s}\alpha_{s} \max_{i\in I(v_s)}n_i\cdot f(v_s) =\sum_{s}\alpha_{s}D^+_fV(v_s). 
\end{split}
\end{equation*}
Since $\alpha_s>0$ and $D^+_fV(v_s)\leq 0$ for every $s$, then $D^+_fV(\bar{x})=0$ only if $D^+_fV(v_s)=0$ for all the vertices $v_s\in S_{\bar{x}}$, which happens only if $v_s\in \cO$ for every vertex $v_s\in S_{\bar{x}}$. For this case, since the set $\cO$ is affine, then $\bar{x}\in \cO$. To sum up, for any $x\in \partial X$, if $D^+_fV(x)=0$, then $x\in \cO$. Since the vector field on $\partial (\lambda X)$ represents $\lambda$-scaled vectors of the vector field on $\partial X$ for all $\lambda\in (0,1)$, it can be easily shown that for any $x\in X$, if $D^+_fV(x)=0$, then $x\in \cO$, i.e. $\{x\in X~:~D^+_fV(x)= 0\}\subset \cO$. Recall that $D^+_fV(x)\leq 0$ for all $x\in X$. By LaSalle's Invariance Principle, we know that the trajectories $\phi(x_0,t,u_p)$ tend to $\{x\in X~:~D^+_fV(x)= 0\}\subset \cO$ as $t\rightarrow \infty$.

Combining the two parts above, for any $x_0\in X^{\circ}$, $\phi(x_0,t,u_p)$ eventually tends to $\cO$ through $X^{\circ}$. %~~~~~~~~~~~~~\QED     

\end{document}